\documentclass[12pt]{article}
\usepackage{fullpage,setspace}
\usepackage{amssymb, amsthm, amsmath}
\usepackage{graphicx}
\usepackage[authoryear]{natbib}
\usepackage{bm}
\usepackage{times}
\usepackage{setspace}
\usepackage{wrapfig}
\usepackage{epstopdf}
\usepackage{subcaption}
\usepackage{caption}
\usepackage{multirow}
\usepackage{verbatim}
\usepackage{url}

\newcommand{\bbeta}{ \mbox{\boldmath $\beta$}}

\newcommand{\bphi}{ \mbox{\boldmath $\phi$}}

\newcommand{\E}{\mbox{E}}
\newcommand{\Var}{\mbox{Var}}

\newcommand{\bA}{ \mbox{\bf A}}

\newcommand{\bP}{ \mbox{\bf P}}

\newcommand{\bX}{ \mbox{\bf X}}

\newcommand{\bs}{ \mbox{\bf s}}

\newcommand{\bu}{ \mbox{\bf u}}

\newcommand{\bC}{ \mbox{\bf C}}
\newcommand{\bg}{ \mbox{\bf g}}

\newcommand{\bI}{ \mbox{\bf I}}

\newcommand{\bR}{ \mbox{\bf R}}

\newcommand{\indep}{\stackrel{indep}{\sim}}

\newcommand{\calR}{{\cal R}}

\newcommand{\beq}{ \begin{equation}}
\newcommand{\eeq}{ \end{equation}}
\newcommand{\beqn}{ \begin{eqnarray}}
\newcommand{\eeqn}{ \end{eqnarray}}

\begin{document} 

\begin{center}
{\Large The R2D2 prior for generalized linear mixed models}\\\vspace{6pt}
{\large Eric Yanchenko\footnote[1]{North Carolina State University}, Howard D. Bondell\footnote[2]{ University of Melbourne} and Brian J. Reich$^1$}\\
\today
\end{center}

\begin{abstract}
\noindent
In Bayesian analysis, the selection of a prior distribution is typically done by considering each parameter in the model. While this can be convenient, in many scenarios it may be desirable to place a prior on a summary measure of the model instead. In this work, we propose a prior on the model fit, as measured by a Bayesian coefficient of determination ($R^2)$, which then induces a prior on the individual parameters. We achieve this by placing a beta prior on $R^2$ and then deriving the induced prior on the global variance parameter for generalized linear mixed models. We derive closed-form expressions in many scenarios and present several approximation strategies when an analytic form is not possible and/or to allow for easier computation. In these situations, we suggest approximating the prior by using a generalized beta prime distribution and provide a simple default prior construction scheme. This approach is quite flexible and can be easily implemented in standard Bayesian software. Lastly, we demonstrate the performance of the method on simulated and real-world data, where the method particularly shines in high-dimensional settings, as well as modeling random effects.

\end{abstract}

\section{Introduction}\label{s:intro}

As computing power has increased and become more accessible, Bayesian inference has risen to prominence.  Researchers are now free to consider complex models with many parameters.  An advantage of the Bayesian approach is that it can incorporate prior domain knowledge about some parameters to reduce uncertainty.  
In the absence of such information, we might select {\it vague} prior distributions, i.e., prior distributions with large variance. This, however, can lead to some unintended consequences such as Lindley's paradox \citep{lindley1957}. Vague prior distributions can also lead to poor estimates when the number of parameters is large relative to the sample size. This has led to the recent development of  shrinkage prior distributions \citep{george1993, rovckova2018, park2008, hans2009, carvalho2010, bhadra2017, bhattacharya2015, zhang2022bayesian}.

Typically, prior distributions are selected for individual parameters based on domain expertise and/or using a general paradigm, e.g., shrinkage priors. There are situations, however, where researchers may have meaningful prior information on the model in general as opposed to specific regression coefficients. For example, consider genetic association studies \citep[e.g.,][]{lewis2012introduction} where scientists search for the genes that contribute to a specific disease. There may be good understanding of how much genes affect the disease, but little information about which genes are relevant. In this case, it may make more sense to pick a prior for the overall model fit that then induces prior distributions on the parameters. There has been some previous work towards this end. \cite{hodges2001counting} use a flat prior distribution on the degrees of freedom in a Gaussian mixed effects model. \cite{simpson2017penalising} introduce a paradigm that penalizes the complexity of the model as measured by the Kullback-Liebler (KL) divergence between the null and fitted model. This method places a prior on this KL divergence, thus shrinking the entire model instead of the individual parameters.  \cite{fuglstad2020intuitive} present a user-friendly approach to prior construction by utilizing prior beliefs to apportion the overall variance between different random effect components. The authors construct a joint prior distribution which considers the entire model structure. For Gaussian linear regression, \cite{zhang2022bayesian} place a prior on the model fit as measured by the coefficient of determination, $R^2$. The authors first derive a Bayesian $R^2$ and show that the prior $R^2\sim\mbox{Beta}(a,b)$ yields a beta prime prior on the total variance of the regression parameters which is then distributed to each individual parameter through a Dirichlet Decomposition. For sparse high-dimensional regression problems, certain $R^2$ prior choices and a Dirichlet decomposition give posterior consistency. This method is advantageous because $R^2$ is an intuitive measure of model fit and it has excellent shrinkage properties.

In this work, we consider a prior on a summary of model fit by proposing a beta prior on \cite{zhang2022bayesian}'s definition of $R^2$ for generalized linear mixed models. This extends \cite{zhang2022bayesian} beyond linear regression to allow for non-Gaussian responses and random effects. We derive closed-form expressions in multiple scenarios for the prior of the global variance parameter that induces a beta prior on $R^2$. We also present several approximation strategies when an analytic prior distribution is not possible. The main approach we suggest approximates the prior by a {\it generalized beta prime} (GBP) distribution. This distribution is quite flexible as it can achieve boundedness at the origin as well as a heavy tail \citep{Perez::2007aa}. The scaled beta prime distribution, a special case of the GBP, has also previously been used as a prior for the variance of the regression coefficients \citep{klein2021bayesian, Bai::2021aa}. Our method, like \cite{zhang2022bayesian}, differs from these previous approaches in that we place a GBP prior on the global variance which is then further decomposed in the hierarchy to the individual regression parameters. Our approach also provides an intuitive way to construct informative prior distributions as well as an automatic approach. The proposed methods can be applied using the {\tt r2d2glmm} package available on GitHub at \url{https://github.com/eyanchenko/r2d2glmm}.

The remainder of the paper proceeds as follows.
In Section \ref{s:glmm}, we describe the generalized linear mixed model framework and present several specific examples. In Section \ref{s:R2_SNR}, we precisely define a Bayesian $R^2$ and show how the model-level prior induces prior distributions for the individual model prior parameters. We also present the prior distributions for several specific regression models as well as approximation techniques when a closed-form solution cannot be found. Section \ref{s:app} applies the proposed method to real-world data and Section \ref{s:discussion} concludes with recommendations for default use and next steps.

\section{Generalized linear mixed models}\label{s:glmm}

For notational simplicity, we follow \cite{simpson2017penalising} and specify our model for a generalized linear mixed model (GLMM), although the ideas presented here can generalize to other settings.  For observations $i\in\{1,\dots,n\}$, let $Y_i$ be the response, $\bX_i = (X_{i1},\dots,X_{ip})$ 
be the explanatory variables and $\bbeta= (\beta_1,\dots,\beta_p)^T$ be the corresponding fixed effects. We standardize the explanatory variables such that each column of $\bX$ has mean zero and variance one.  We also assume that there are $q$ types of random effects, $\bu_k$, $k\in\{1,\dots,q\}$ where $\bu_k= (u_{k1},...,u_{kL_k})^T$ has $L_k$ levels. We let ${\bf g}_{i}=(g_{i1},\dots,g_{iq})^T$ for $i\in\{1,\dots,n\}$ be membership vectors such that $g_{ik}$ is the level of random effect $k$ for observation $i$ and where mixed-membership is excluded. The fixed and random effects prior distributions are assumed to be independent and are related to the response via the linear predictor 
\begin{equation}\label{e:linear_predictor}
  \eta_i = \beta_0 + \bX_i\bbeta + \sum_{k=1}^q u_{kg_{ik}}
\end{equation}
where $\beta_0$ is the intercept. The responses are assumed to be conditionally independent given the linear predictor and follow density function $Y_i|\eta_i,\theta\sim f(y|\eta_i, \theta)$, where $\theta$ is an additional parameter in the likelihood function (see examples below).

The model for the fixed and random effects is $\beta_j|\phi_j,W\indep\mbox{Normal}(0, \phi_j W)$ and $\bu_k|\phi_{p+k},W\indep\mbox{Normal}(0, \phi_{p+k}W \bI_{L_k})$ where $W>0$ controls the overall variance of the linear predictor (not the response) and $\phi_j\ge0$ satisfy  $\sum_{j=1}^{p+q}\phi_{j}=1$ and apportion the variance to the different model components. Thus, $W$ may be interpreted as the total amount of variation in the fixed and random effects, or as a transformation of the total variation of the mean function. In the latter case, the interpretation depends on the link function. Moreover, large values of $W$ encode a model with greater flexibility since large variance in the mean function means that the model can capture more trends in the data. In the limit as $W\to0$, conversely, we are reduced to the intercept-only model. This interpretation will be important later in this work when we treat the placement of a large prior mass on $W$ near zero as ``penalizing" towards the null (intercept-only) model. Additionally, notice that the fixed and random effects are modeled similarly, i.e., with a random variance. Even so, we maintain their differing interpretations. Specifically, if we are interested in effect estimates themselves, then we treat this effect as ``fixed,'' but if our interest lies in the underlying population of the effect, then it is treated as ``random" \citep{searle2009variance}. Following this interpretation, we are most interested in the estimates of $\bbeta$ and $\phi_j W$ for $j=p+1,\dots,p+q$.

The prior distribution of $R^2$ relies on the distribution of $\eta_i$. For the majority of this work, we assume
\begin{equation}\label{eq:eta_dist}
    \eta_i|\beta_0,W \sim \mbox{Normal}(\beta_0, W).
\end{equation} 
We derive this result in the Supplemental Materials whether $\bX_i$ is treated as fixed or random. If we treat $\bX_i$ as random, then $\eta_i$ will be approximately normal for moderate $p$ by the Central Limit Theorem. On the other hand, if we consider $\eta_i$ conditional on $\bX_i$, then the distribution of $\eta_i$ is exactly normal where the variance is different for each $i$ but the average variance is $W$ due to $\bX$'s standardization. For either case, we stress that the prior distribution of $\eta_i$ is independent of the explanatory variables, resulting in a prior that does not depend on $\bX$, similar to the PC prior \citep{simpson2017penalising}. Alternative distributions are discussed in Sections \ref{s:QMC} - \ref{s:gbp} but the normality of $\eta_i$ is assumed for all experiments.


\subsection{Variance decomposition of the linear predictor}

The variance parameters $\bphi = (\phi_1,...,\phi_{p+q})$ determine the relative variance of each component of the model and are restricted to sum to one.  These parameters could be fixed, or given prior distributions to add flexibility to the variance decomposition.   In the most general case we can assign these parameters a Dirichlet distribution, $\bphi\sim\mbox{Dirichlet}(\xi_1,...,\xi_{p+q})$. Often times we will take $\xi_1=\cdots=\xi_{p+q}\equiv \xi_0$.  The concentration parameter $\xi_0>0$ controls the variation of the prior distribution with large $\xi_0$ encouraging all the variance components to be roughly equal to $1/(p+q)$, and small $\xi_0$ reflecting prior uncertainty in the variance components.  In some cases, the effects will be grouped and the variance across groups will be decomposed using a Dirichlet prior, e.g., all fixed effects assumed to have the same variance.  These ideas are illustrated through examples below.

\subsection{Examples}\label{s:examples}
To help fix ideas, we present a few specific examples of this prior construction.

\paragraph{Example 1: Gaussian linear regression model:} In the linear regression setting with no random effects, the linear predictor is simply $$\eta_i=\beta_0+\bX_i\bbeta$$ and we have $Y_i|\eta_i,\sigma^2\sim\mbox{Normal}(\eta_i,\sigma^2)$ so that $\theta=\sigma^2$ is the error variance. We then take $\beta_j|\phi_j,W\sim\mbox{Normal}(0,\phi_j W)$ for $j=1,\dots,p$. \cite{zhang2022bayesian} study the theoretical properties of this approach for various prior distributions on $W$ and $\bphi$. In general, this is a global-local shrinkage prior which has been studied in various contexts \citep[e.g.,][]{carvalho2010, polson2012bb, Polson2012shrinkage, bhattacharya2015, zhang2018}.

\paragraph{Example 2: Poisson regression with two-way random effects:} For a mixed effects model with two-way (non-interacting) random effects, the linear predictor is $$\eta_{i}=\beta_0+\bX_i\bbeta + u_{1g_{i1}} + u_{2g_{i2}},$$ and $Y_i|\eta_i\sim\mbox{Poisson}\{\exp(\eta_i)\}$. The membership vectors $g_{i1}$ and $g_{i2}$ indicate the level assigned to observation $i$ for random effects type one and two, respectively. The variance weights given to the fixed and random effects are determined by the Dirichlet parameter $\bphi$. For example, to allow each fixed effect to have a different variance, we might take $\bphi\sim\mbox{Dirichlet}(\xi_1,\dots,\xi_{p+2})$ where $\xi_k$ are fixed hyperparameters; on the other hand, for each fixed effect to have the same variance, we might take $\bphi\sim\mbox{Dirichlet}(\xi_1,\xi_2,\xi_3)$ and then let $\beta_j|\phi_1,W\sim\mbox{Normal}(0,\tfrac1p\phi_1W)$ for $j=1,\dots,p$ and $\bu_k\sim \mbox{Normal}(0, \phi_kW\bI_{L_k})$ for $k=2,3$.

\paragraph{Example 3: Weibull model:} Survival analysis often uses a Weibull model. For simplicity, we consider uncensored data but this could be extended to censored data. Let there be a single random effect so that the linear predictor is
$$
    \eta_i = \beta_0 + \bX_i\bbeta + u_{g_i}
$$
for membership vector $g_i\in\{1.\dots,L\}$.
If $Y_i$ is the survival time, then the model is $Y_i|\eta_i,\theta\sim\mbox{Weibull}(e^{\eta_i},\theta)$ for shape parameter $\theta$. If we assume that the fixed effects have equal variance, then $\bbeta|\phi_1,W\sim\mbox{Normal}(0,\tfrac1p\phi_1W\bI_{p})$ and $\bu|\phi_2,W\sim\mbox{Normal}(0,\phi_2W\bI_{L})$ where $\boldsymbol\phi\sim\mbox{Dirichlet}(\xi_0,\xi_0)$.

\paragraph{Example 4: Generalized linear regression with spatial random effects:} Consider the scenario where we observe data from $L$ spatial clusters (e.g., cities or villages) at spatial locations $\bs_1,\dots,\bs_L\in\calR^2$. Then let $Y_i$ be the response from location $\bs_{g_i}\in\calR^2$ where $g_i\in\{1,\dots,L\}$ is the cluster indicator. Spatial generalized linear models account for correlation between observations at nearby locations by adding spatially-correlated random effects \citep[e.g.,][]{diggle1998model}. Let $u_{g_i}$ be the Gaussian random effect for cluster $g_i$.  The linear predictor is then $\eta_i = \beta_0 + \bX_i\bbeta + u_{g_i}$.  A stationary and isotropic model assumes $\mbox{E}(u_i)=0$ and $\mbox{Var}(u_i)=\sigma_u^2$ for all $i$ and $\mbox{Cor}(u_i,u_j)=C(d_{ij})$, where $C$ is a spatial correlation function such as the exponential function $C(d) = \exp(-d/\rho)$ and $d_{ij}$ is the distance between locations $\bs_i$ and $\bs_j$. The covariance structure of the model is determined by the $L\times L$ correlation matrix $\bC$ with $(i,j)$ element $C(d_{ij})$.  The spatial regression model is then in the form of (\ref{e:linear_predictor}) where $\bu|\phi_{p+1},W,\rho \sim\mbox{Normal}(0,\phi_{p+1}W{\bf C})$ and $\sigma^2_u=\phi_{p+1}W$. While the covariance matrix of the random effect is no longer diagonal, the derivation of (\ref{eq:eta_dist}) still holds as the different random effect levels have the same variance and the covariance terms do not appear in the derivation.



\paragraph{Example 5: Generalized additive model:} Non-linear regression models can also be written as (\ref{e:linear_predictor}).  Assume that $p$ explanatory variables, $x_{i1},...,x_{ip}$, are allowed to have a non-linear relationship with the response variable. The generalized additive model \citep[e.g.,][]{hastie2017generalized,klein2021bayesian} is $$\eta_i = \beta_0+\sum_{j=1}^p f_j(x_{ij})$$ for unknown functions $f_1,...,f_p$. A common approach is to model the $f_j$'s using a basis expansion 
$$
f_j(x) = \sum_{l=1}^{L_j}B_{jl}(x)\beta^{(k)}_{l}
$$
where $B_{jl},...,B_{jL_j}$ are basis function, e.g., spline functions and $\bbeta^{(k)}$ are ``grouped" fixed effects.  This model then fits (\ref{e:linear_predictor}) with $\tilde{\bX}=(\tilde{\bX}_1,\dots,\tilde {\bX}_p)$ where $\tilde{\bX}_j\in\mathcal R^{n\times L_j}$ is such that $(\tilde{\bX}_j)_{ik}=B_{jk}(x_{ij})$, and $\bbeta=({\bbeta^{(1)}}^T,\dots,{\bbeta^{(p)}}^T)^T$. Then $\beta^{(j)}_k \sim \mbox{Normal}(0, \tfrac1{L_j}\phi_j W)$ for $j\in\{1,\dots,p\}$ and $k\in\{1,\dots,L_j\}$ such that $\phi_j$ determines the proportion of the variance allocated to the non-linear effect of $x_{ij}$.

\section{Variance Decomposition $R^2$ and the R2D2 prior}\label{s:R2_SNR}

\cite{gelmanf2019r}, \cite{gelman2006data} and \cite{zhang2022bayesian} propose measures of model complexity that we name the {\it Variance Decomposition} $R^2$ (VaDeR).  For the GLMM in Section \ref{s:glmm}, define $\mbox{E}(Y_i|\eta_i)=\mu(\eta_i)$ and $\mbox{Var}(Y_i|\eta_i)=\sigma^2(\eta_i)$ which relates the linear predictor to the response distribution. \cite{gelmanf2019r} use the empirical definition of $R^2$, defined as \begin{equation}\label{e:R2_gelman_finite}
   R_n^2 =  \frac{\mbox{V}\{\mu(\eta_1),...,\mu(\eta_n)|\bX,\bg,\bbeta,\bu\}}{\mbox{V}\{\mu(\eta_1),...,\mu(\eta_n)|\bX,\bg,\bbeta,\bu\}+\mbox{M}\{\sigma^2(\eta_1),...,\sigma^2(\eta_n)|\bX,\bg,\bbeta,\bu\}}
\end{equation}
where M and V are the sample mean and variance operators, respectively.

In (\ref{e:R2_gelman_finite}), $\mbox{V}\{\mu(\eta_1),\dots,\mu(\eta_n)|\bX,\bg,\bbeta,\bu\}$ is the variance of the expectation of future data and $\mbox{M}\{\sigma^2(\eta_1),\dots,\sigma^2(\eta_n)|\bX,\bg,\bbeta,\bu\}$ is the expected variance of future residuals, both conditioned on the explanatory variables, membership vectors and fixed and random effects. Because of this conditioning, \cite{gelmanf2019r} propose $R^2_n$ as an {\it a posteriori} measure of model fit. In principle, however, if the values of $\bX_i$ and $\bg_{i}$ are known but we had yet to observe the responses $Y_i$, then the prior distributions of the fixed and random effects would induce a prior distribution on $R^2_n$.
Then $R^2_n$ is the proportion of variance explained by the model for future data, conditioned on these variables and our prior information for $\bbeta$ and $\bu_k$.

While $R_n^2$ is an intuitive measure of the fit of the model to a particular dataset, for the purpose of setting prior distributions we follow \cite{zhang2022bayesian}. We measure complexity at the population level and use the marginal version of $R^2$ that averages over variation in both the explanatory variables and random effect levels ($\bX$ and ${\bf g}$) as well as parameters ($\bbeta$ and $\bu_k$). The marginal distribution does not depend on $\bX_i$ or ${\bf g}_i$ so the observations are exchangeable. We can then drop the subscript distinguishing them and consider the model for an arbitrary observation $Y$ with $\mbox{E}(Y|\eta)=\mu(\eta)$, $\mbox{Var}(Y|\eta)=\sigma^2(\eta)$ and $\eta|\beta_0,W \sim \mbox{Normal}(\beta_0,W)$ as in (\ref{eq:eta_dist}).  Then $R^2$ becomes
\begin{equation}\label{e:R2_zhang}
   R^2(\beta_0,W) = \frac{\mbox{Var}\{\mu(\eta)\}}{\mbox{Var}(Y)} = \frac{\mbox{Var}\{\mu(\eta)\}}{\mbox{Var}\{\mu(\eta)\}+\mbox{E}\{\sigma^2(\eta)\}}
\end{equation}
where $\mbox{E}\{\sigma^2(\eta)\}$ and $\mbox{Var}\{\mu(\eta)\}$ are summaries of the distribution of $\eta$ and thus depend on parameters $\beta_0$ and $W$. For the sake of simplicity, we suppress the dependence on $(\beta_0,W)$ and write $R^2(\beta_0,W)=R^2$ for the remainder of the paper. 
The Supplemental Materials discusses the relationship between $R^2_n$ and $R^2$ and shows that under general conditions, $R_n^2$ will converge to $R^2$ when both the sample size and number of effective parameters increase. We also include a brief discussion comparing $R^2$ and $R^2_n$ with other measures of model fit for GLMMs \citep[e.g.][]{cox1989analysis, mcfadden1973conditional}. Moreover, we note that the coefficient of determination is not commonly used for GLMMs as a measure of model fit. This could be for several reasons, not least of which being that it is difficult to define and interpret a principled $R^2$ for logistic regression, poisson regression, etc. A major advantage of $R^2$ and $R^2_n$ is that they can easily be extended to GLMMs, while also having intuitive interpretations.

As denoted in (\ref{e:R2_zhang}), the prior distribution of $R^2$ is determined by the joint prior ($\beta_0,W)$.  For Gaussian responses the distribution of $R^2$ is invariant to $\beta_0$, and so to reduce the problem to matching univariate distributions, we parameterize the prior for $(\beta_0,W)$ as the conditional prior for $W|\beta_0$ and marginal prior for $\beta_0\sim\pi_0$.  We then select a prior for $W|\beta_0$ so that $R^2\sim\mbox{Beta}(a,b)$.  By construction, since $R^2\sim\mbox{Beta}(a,b)$ conditioned on any $\beta_0$, $R^2$ also follows a $\mbox{Beta}(a,b)$ marginally over the joint prior for $(\beta_0,W)$ for any marginal prior $\pi_0$. Combined with the Dirichlet prior distribution on the variance proportions, this defines the {\it $R^2$ Dirichlet decomposition prior} (R2D2).  

The Beta$(a,b)$ prior for $R^2$ is our default choice, but in some cases the support of $R^2$ can be restricted to a subspace of $[0,1]$ and a modification is required. Typically, when $W=0$ we also have $\mbox{Var}\{\mu(\eta)\}=0$ and thus $R^2=0$ assuming the distribution of $Y|\eta$ is not degenerate, i.e., $\sigma^2(\eta)>0$. If, however, $\mbox{Var}\{\mu(\eta)\}>0$ when $W=0$, then the lower bound of $R^2$, $R^2_{min}$, is strictly greater than zero (e.g. Poisson regression with offsets in Supplemental Material).  Conversely, for some link functions, $R^2<1$ for all $W$ (e.g., the zero-inflated Poisson model in Supplemental Materials). In general, the upper bound of $R^2$, $R^2_{max}$, is
1 if and only if
$
    \E\{\sigma^2(\eta)\}
    =o\big(\Var\{\mu(\eta)\}\big)
$
as $W\to\infty$. In cases where $R^2_{min}>0$ and/or $R^2_{max}<1$, we use a Beta$(a,b)$ prior distribution for the shifted and scaled $R^2$, denoted ${\tilde R}^2=(R^2-R^2_{min})/(R^2_{max}-R^2_{min})$. This is equivalent to using a {\it four-parameter beta distribution} for the prior where $R^2\sim\mbox{Beta}(a,b,R^2_{min},R^2_{max})$ has density function
$$
  \pi(r^2)
  =\frac{(r^2-R^2_{min})^{a-1}(R^2_{max}-r^2)^{b-1}}{(R^2_{max}-R^2_{min})^{a+b-1}B(a,b)},\ R^2_{min}\leq r^2\leq R^2_{max}.
$$
In most cases, $R^2_{min}=0$ and $R^2_{max}=1$ so unless otherwise noted we simply denote the prior as $R^2\sim\mbox{Beta}(a,b)$. 

\subsection{Special cases with exact expressions}\label{s:specialcases}

Below we derive the expressions for the prior distribution for $W$ in several special cases where the exact prior distribution is available.

\begin{figure}
    \centering
    \includegraphics[width=\textwidth]{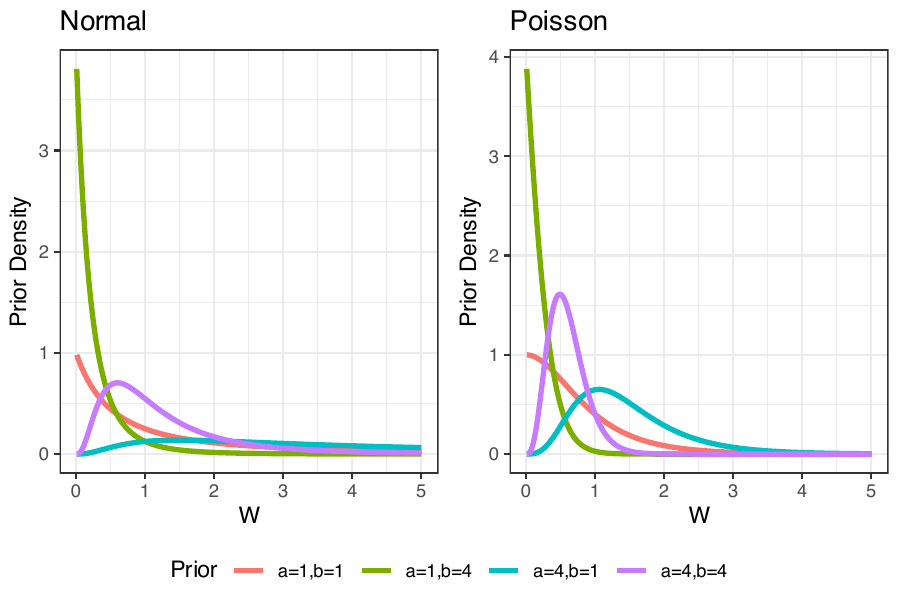}
    \caption{Plot of the prior distribution of $W$ to induce $R^2\sim\mbox{Beta}(a,b)$ with $\beta_0=0$. The title of each panel corresponds to the response model. The normal case takes $\sigma^2=1$.}
    \label{fig:W_prior}
\end{figure}

\paragraph{Location-scale  models:}
The location-scale model is $Y_i = \eta_i+\sigma\epsilon_i$, where the errors $\epsilon_i$ have mean zero and variance one. 
Then $\mu(\eta) = \eta$ and $\sigma^2(\eta) = \sigma^2$ and thus $R^2 = W/(W+\sigma^2)$.  Assuming $R^2$ follows a Beta($a,b$) and $\sigma=1$ (or more generally that $\sigma^2$ appears in the prior variance, $\beta_j|\sigma^2,\phi_j,W\sim\mbox{Normal}(0,\sigma^2\phi_jW)$), \cite{zhang2022bayesian} show that the induced prior on $W$ is a Beta Prime distribution, denoted $W\sim\mbox{BP}(a,b)$ with density  function
\begin{equation}
    \pi(w)
    =\frac{1}{B(a,b)}\frac{w^{a-1}}{(1+w)^{a+b}},\ w\geq0,
\end{equation}
where $B(\cdot,\cdot)$ denotes the Beta function.  In the left panel of Figure \ref{fig:W_prior}, we plot $\pi(w)$ for various values of $a$ and $b$, and we can see that the BP prior distribution for $W$ has heavier tails when the expected $R^2$ is large $(a>b)$ versus small $(a<b)$.

For $\sigma^2\ne 1$, and not included in the prior variance, i.e., $\beta_j|\phi_j,W\sim\mbox{Normal}(0,\phi_j W)$, the induced prior distribution for $W$ is a {\it Generalized Beta Prime} (GBP) distribution, $W|\sigma^2\sim\mbox{GBP}(a,b,1,\sigma^2)$. The GBP distribution can be obtained via a transformation of a BP random variable, i.e., if $V\sim\mbox{BP}(a,b)$ then $W=dV^{1/c}\sim\mbox{GBP}(a,b,c,d)$ and has density function
\begin{equation}\label{e:GBP}
    \pi(w;a,b,c,d)
    =\frac{c\left(\frac{w}{d}\right)^{a c-1}\left(1+\left(\frac wd\right)^c\right)^{-a-b}}{dB(a,b)},\ w\geq0
\end{equation} 
for $a,b,c,d>0$.  The GBP reduces to the BP if $c=d=1$. 

We note a few properties of the GBP distribution. The behavior at the origin is controlled by the value of $ac$, with
$$
    \lim_{w\rightarrow 0}\pi(w;a,b,c,d) = 
    \begin{cases}
      \infty & ac < 1\\
      \frac{c}{B(a,v)d} & ac=1\\
      0 & ac > 1
    \end{cases}.
$$
The tail behaviour is controlled by $bc$ with valid mean if only if $bc>1$.  Also, for any model with $W\sim\mbox{GBP}(a,b,c,d)$ for the overall variance, then the standard deviation has prior distribution $W^{1/2}\sim\mbox{GBP}(a,b,2c,d^{1/2})$. As another special case of the GBP, if $a=1/2$, $b=\nu/2$, $c=2$ and $d=\sqrt{\nu\sigma^2}$, then $W$ is distributed as a half-$t$ distribution with $\nu$ degrees of freedom and scale $\sigma^2$. Specifically, if $W\sim\mbox{GBP}(\frac12,\frac12, 1, \sigma^2)$, then $\sqrt{W}$ follows a half-Cauchy distribution with scale $\sigma$ as in \cite{Gelman2006::cauchy}.


\paragraph{Poisson regression:}
The Poisson regression model is $Y|\eta \sim\mbox{Poisson}(e^\eta)$ and thus $\mu(\eta)=\sigma^2(\eta)=e^\eta$.  Since $\eta|\beta_0,W\sim\mbox{Normal}(\beta_0,W)$, $e^\eta|\beta_0,W\sim\mbox{LogNormal}(\beta_0,W)$, and thus
\begin{equation}\label{e:PoissonR2}
    R^2
    =\frac{e^{W}-1}{e^{W}-1+e^{-\beta_0-\frac12W}}.
\end{equation}
$R^2\sim\mbox{Beta}(a,b)$ induces (see Supplemental Materials) the prior for $W$ with density
\begin{equation}\label{deriv:PoissW}
    \pi(w| \beta_0;a,b)
    =\frac{1}{B(a,b)} \frac{(e^{w}-1)^{a-1} e^{-b(\beta_0+w/2)}(3e^{w}-1)}{2(e^{w}-1+e^{-\beta_0-w/2})^{a+b}},\ w\geq0.
\end{equation}
We plot this distribution in Figure \ref{fig:W_prior}
and show that shape of the prior looks very similar to that of the location-scale case. We do note, however, that the prior for $W$ has exponential-decaying tails on the scale of $\mbox{E}(Y|\eta)=e^\eta$ as seen in (\ref{deriv:PoissW}). But, on the scale of $\log\{\mbox{E}(Y|\eta)\}=\eta$, which is the same scale as $\bbeta$ and $\bu$, the prior has polynomial-decaying tails. The value of the prior at 0 is $\infty$ if $a<1$, $be^{\beta_0}$ if $a=1$ and $0$ if $a>1$.

In the Supplemental Materials, we also include the exact prior distributions for: Poisson with offsets, negative binomial, zero-inflated Poisson, and the Weibull model.

\subsection{Approximate Methods}
In some cases, e.g., logistic regression, a closed-form expression for VaDeR is not available, so in this section we discuss alternatives.

\subsubsection{Linear approximation}\label{s:deltamethod}
The two components we must compute for VaDeR are $\mbox{Var}\{\mu(\eta)\}$ and $\mbox{E}\{\sigma^2(\eta)\}$. The simplest approach to approximate these is with a linear approximation. Applying a first-order Taylor series approximation of $\mu(\eta)$ and $\sigma^2(\eta)$ around $\beta_0$ gives
\begin{equation}
    \mbox{Var}\{\mu(\eta)\}
    \approx \{\mu'(\beta_0)\}^2W
\mbox{\ \ \ and \ \ \ }
    \mbox{E}\{\sigma^2(\eta)\}
    \approx \sigma^2(\beta_0).
\end{equation}
Then denoting $s^2(\beta_0) = \sigma^2(\beta_0)/\{\mu'(\beta_0)\}^2$ we have
\begin{equation}\label{eq:delta}
    R^2
    \approx \frac{W}{W+s^2(\beta_0)}.
\end{equation}
If $R^2\sim\mbox{Beta}(a,b)$, the resulting prior for $W$ is $W|\beta_0\sim\mbox{GBP}(a,b,1,s^2(\beta_0))$. This result does not require any distributional assumptions about $\eta_i$ other than a finite mean and variance after transformation by $\mu(\cdot)$ and $\sigma^2(\cdot)$. Additionally, the computational burden is essentially zero.

\subsubsection{Quasi-Monte Carlo (QMC)}\label{s:QMC}
As we will show, in many cases the linear approximation dose a poor job approximation the true $R^2$ distribution. Therefore, we must turn to other methods. Since finding $R^2$ reduces to computing complicated integrals, we can use integral approximation techniques, like quasi-Monte Carlo  \citep[QMC; e.g.,][]{Morokoff1995}. In usual Monte Carlo integration, the integral of interest is approximated by summing over a randomly generated sample of points.
QMC is similar except that the points are selected deterministically. To construct the R2D2 prior, we approximate
\begin{equation}\label{eq:qmc1}
    \E\{\mu(\eta)^m\}
    \approx {\tilde \mu}_m(W|\beta_0) = \frac1{K-1}\sum_{i=1}^{K-1} \mu(\beta_0+z_i\sqrt{W})^m
\end{equation}
and
\begin{equation}\label{eq:qmc2}
    \E\{\sigma^2(\eta)\}
    \approx {\tilde \sigma}^2(W|\beta_0)=\frac1{K-1}\sum_{i=1}^{K-1} \sigma^2(\beta_0+z_i\sqrt{W})
\end{equation}
where $z_i$ is the $i/K$ quantile of a standard normal distribution and $m=1,2$.  This gives an approximation of $R^2$ for a given $\beta_0$ and $W$, which we denote by
\begin{equation}\label{eq:qmc3}
    {\tilde R}^2(W|\beta_0)
    \approx\frac{{\tilde \mu}_2(W|\beta_0)-{\tilde \mu}_1^2(W|\beta_0)}
    {{\tilde \mu}_2(W|\beta_0)-{\tilde \mu}_1^2(W|\beta_0) + {\tilde \sigma}^2(W|\beta_0)}.
\end{equation} 

Assuming $R^2\sim\mbox{Beta}(a,b)$, then the prior for $W$ is
\begin{equation}
    \pi(w|\beta_0;a,b)
    =\frac1{B(a,b)}\{\tilde R^2(w|\beta_0)\}^{a-1}\{1-\tilde R^2(w|\beta_0)\}^{b-1} \left|\frac{d \tilde R^2(w|\beta_0)}{dw}\right|,\ w\geq 0.
\end{equation}
Since this cannot be represented with elementary operations, in practice, we take a numerical derivative to evaluate the prior at a given value, and the entire calculation is completed virtually immediately.

The results in (\ref{eq:qmc1}) and (\ref{eq:qmc2}) make use of the normality of $\eta_i$ from (\ref{eq:eta_dist}). The QMC procedure can be modified to account for non-normal $\eta_i$. Let $\eta\sim F(\eta|\beta_0,W)$ for distribution function $F(\eta|\beta_0,W)$. Then we approximate
$$
    \E\{\mu(\eta)^m\}
    \approx {\tilde \mu}_m(W|\beta_0) = \frac1{K-1}\sum_{i=1}^{K-1} \mu\{q_i(\beta_0,W)\}^m
$$
where $q_i(\beta_0,W)$ is the $i/K$ quantile of $F(\eta|\beta_0,W)$. A similar result holds for approximating $\mbox{E}\{\sigma^2(\eta)\}$ which then leads to an analogous result to (\ref{eq:qmc3}). In practice, $F(\eta|\beta_0,W)$ can be derived analytically if the distribution of $\bX_i$ is known. A more general strategy is to average over the empirical distribution of $\bX$ giving a mixture of normal distributions for $F(\eta|\beta_0,W)$.

\subsubsection{Generalized beta prime approximation}\label{s:gbp}
The GBP distribution provides an exact solution for the location-scale model in Section \ref{s:specialcases}, and an approximate solution for the linear approximation in Section \ref{s:deltamethod}.  The prior $W\sim \mbox{GBP}(a,b,c,d)$ also induces the exact  $R^2\sim\mbox{Beta}(a,b)$ prior distribution for any model with link functions $  \mbox{Var}\{\mu(\eta)\}=W^c$ and $\mbox{E}\{\sigma^2(\eta)\}=d^c$. The GBP will not give an exact solution in all cases, but it is a flexible four-parameter model which may often provide a reasonable approximation. Therefore, a general approximation strategy is to find the values of $(a^*, b^*, c^*, d^*)$ so that the prior $W\sim \mbox{GBP}(a^*, b^*, c^*, d^*)$ gives an approximate $\mbox{Beta}(a,b)$ distribution for $R^2$.

The optimal values of $(a^*, b^*, c^*, d^*)$ depend on $\mu(\cdot)$ and $\sigma^2(\cdot)$ as well as $\beta_0$, $a$ and $b$.  For given link functions and parameters, let $W\sim \pi(w)$ be the distribution that gives exactly $R^2\sim\mbox{Beta}(a,b)$. The GBP parameters are then set to minimize the Pearson $\chi^2$-divergence \citep{renyi1961} between the true and approximated PDFs since this metric enforces a close fit at both the origin and in the tails. We found that minimizing this quantity alone, however, led to unstable solutions, i.e., the surface being maximized over is ``flat." This means that vastly different values of $(a^*,b^*,c^*,d^*)$ may lead to GBP distributions that yield roughly the same approximation of $\pi(w)$. Thus, we also add a regularization term to shrink the prior towards a $\mbox{GBP}(a,b,1,1)$ distribution. We regularize toward this distribution because it gives the exact solution in the location-scale case and can be considered the baseline distribution. This results in the following optimization problem:
\begin{multline}\label{eq:opt}
    (a^*,b^*,c^*,d^*)
    =\underset{\alpha,\beta,c,d}{\operatorname{argmin}}\int_{0}^\infty \left\{\frac{f_{GBP}(w;\alpha,\beta,c,d)-\pi(w)}{\pi(w)}\right\}^2\pi(w)\ dw \\+ \lambda \{(\alpha-a)^2+(\beta-b)^2+(c-1)^2+(d-1)^2\},
\end{multline}
where $\lambda>0$ is a tuning parameter. A larger value of $\lambda$ yields a more stable solution but with a worse fit whereas a smaller value of $\lambda$ yields a better fit but with more instability. We found that $\lambda=\frac14$ gives a good balance between fit and stability. In practice, the integral is approximated by a sum and $\pi(w)$ is approximated using QMC as in Section \ref{s:QMC}, if necessary.  Since the GBP approximation may depend on the QMC procedure which can be modified to allow for non-normality in $\eta$, the GBP approach can similarly be adapted to allow for any distribution of $\eta$.

A major advantage of the GBP prior is that it can be easily implemented in standard Bayesian software such as {\tt JAGS} \citep{plummer2016} or {\tt Stan} \citep{carpenter2017}. Because the exact prior distributions found in Section \ref{s:specialcases}, as well as the resulting distributions from the QMC procedure in Section \ref{s:QMC} do not easily allow for Gibbs sampling, these priors would be difficult for a practitioner to implement. By finding the GBP approximation, however, the R2D2 prior can be easily coded in {\tt JAGS} and {\tt Stan}. Example code is available as a vignette in our R package on GitHub. To specify the prior in these packages, we use the relationship that if $R^2\sim\mbox{Beta}(a,b)$ and $W=d\{R^2/(1-R^2)\}^{1/c}$, then $W\sim\mbox{GBP}(a,b,c,d)$. Because of these features, we recommend this method for general use even in cases when the exact expression is available, and will be the method we consider in the simulation studies.

Since the GBP approximation depends on $\beta_0$ (and $\theta$), this approximation should be updated with the unknown parameter $\beta_0$. The calculation in \eqref{eq:opt} takes about half a second, however, so this would be overly time prohibitive to compute during every MCMC iteration. Instead, we find the GBP approximation once at the beginning of the analysis at $\hat\beta_0=g(\sum_{i=1}^nY_i/n)$ for link function $g(\cdot)$. 
Thus, to induce $R^2\sim\mbox{Beta}(a,b)$, the first step is to find $(a^*,b^*,c^*,d^*)$ as in (\ref{eq:opt}) at $\hat\beta_0$ (and $\hat\theta_{MLE}$, if necessary, the maximum likelihood estimate of the dispersion parameter). After determining ($a^*,b^*,c^*,d^*)$, $\beta_0$ (and $\theta$) are treated as unknown parameters in the subsequent Bayesian analysis. In the Supplemental Materials, we report the GBP approximations for various values of $(a,b)$ and different models. In most cases, the best fitting $a^*$ and $b^*$ values are not close to $(a,b)$ which demonstrates the need for this approximation.

\begin{figure}
    \centering
    {\includegraphics[width = \textwidth]{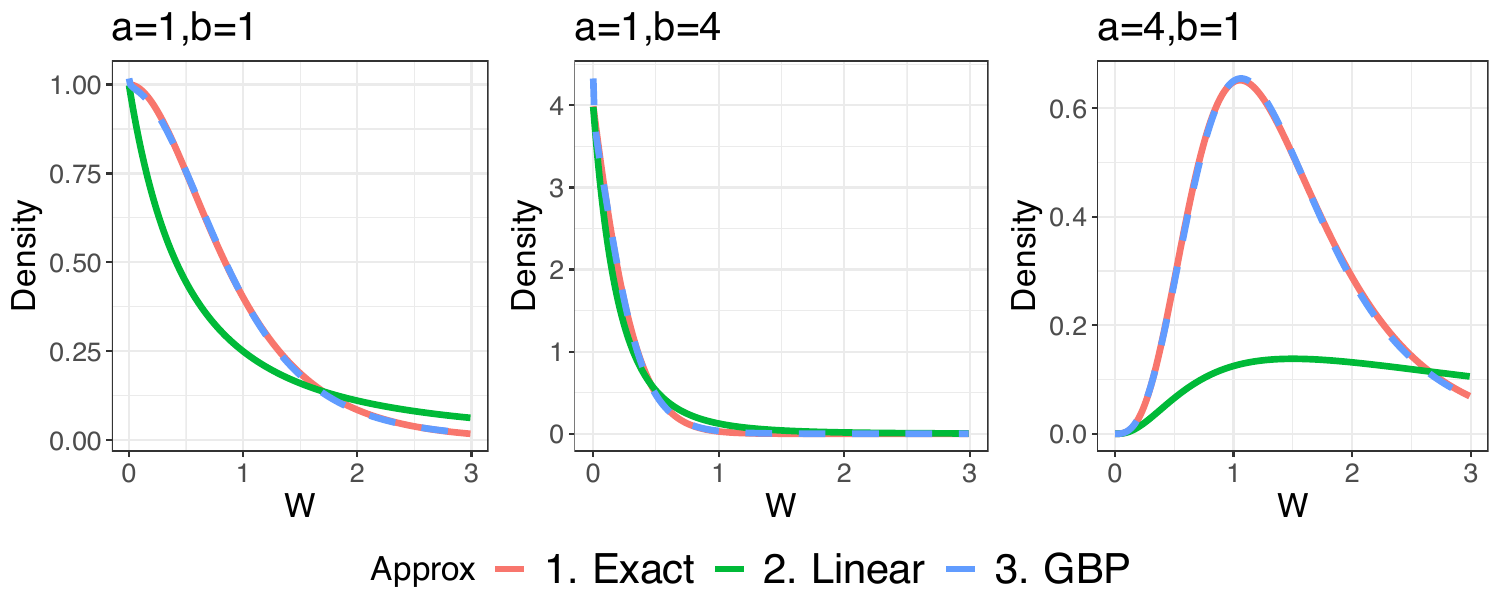}}
    \caption{Comparison of different approximation methods for the Poisson regression model with $\beta_0=0$. The title of each pane indicates the induced prior distribution on $R^2$.}
    \label{fig:pdf_compare}
\end{figure}

Figure \ref{fig:pdf_compare} compares the linear and GBP approximations with the true distribution for the Poisson model (Section \ref{s:specialcases}). The GBP is nearly a perfect match to the true distribution for each prior. The linear approximation is reasonable when $a=1,b=4$, but very poor when $a=4,b=1$. This example shows that the GBP is a very good approximation to the true prior distribution of $W$.

In the Supplemental Materials, we conduct a simulation study to compare the R2D2 prior with a standard vague prior, the Horseshoe prior \citep{carvalho2009handling}, and the PC prior \citep{simpson2017penalising} while also comparing different combinations of $(a,b)$. The proposed method performs favorably across all settings and particularly well in the high-dimensional regression setting. Indeed, we empirically find that prior distributions with large prior mass near $R^2=0$ yield good shrinkage properties.

\section{Real data analysis}\label{s:app}

The proposed method is now applied to two real-world data sets. The first analysis is focused on inference, while the second looks at prediction in the high-dimensional setting.

\subsection{Malaria data}

We now analyze the {\tt gambia} data set \citep{thomson1999predicting} from the {\tt geoR} package \citep{ribeiro2007geor} in {\tt R} to demonstrate the use of the R2D2 prior in practice. We also consider PC and vague prior distributions. There are $n=2035$ children in this data set with binary response variable $Y_i$ which equals 1 if child $i$ tested positive for malaria and 0 otherwise. There are $p=5$ explanatory variables including age, indicator of using a bed net, indicator of whether the bed net is treated, ``greenness" of village and indicator of a health center in the area. These variables are standardized to have mean zero and variance one. There are also the $L=65$ villages where each child lived, along with the spatial location of each village. 

\begin{figure}
    \centering
    \includegraphics[width=\textwidth]{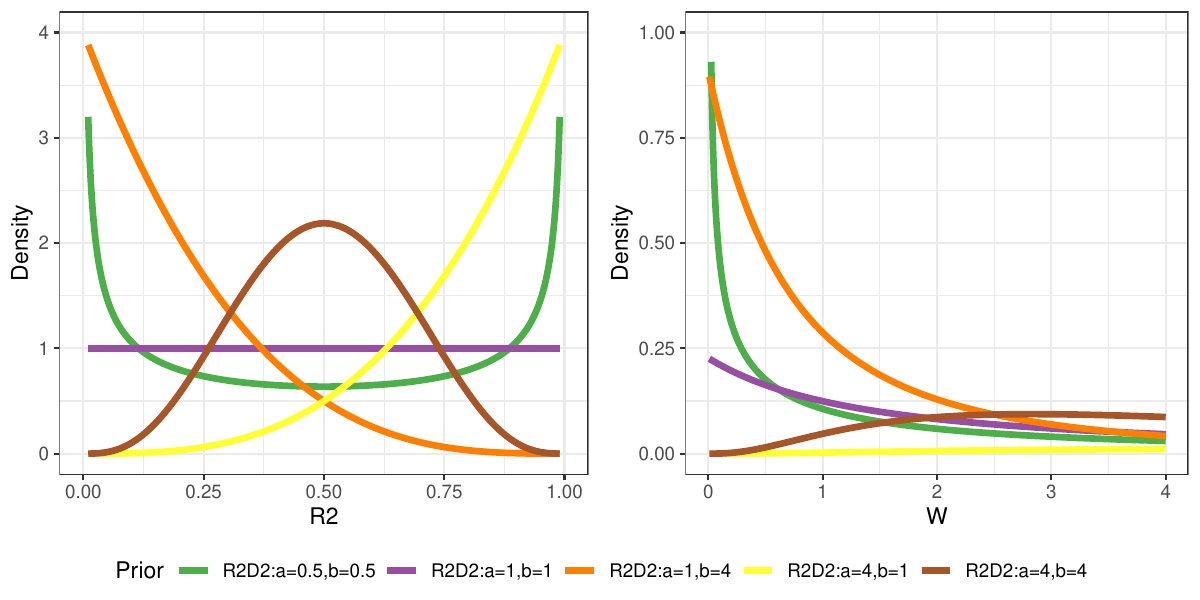}
    \caption{Prior $R^2$ and global variance parameter for R2D2 prior for {\tt Gambia} data.}
    \label{fig:gamb.rand.prior}
\end{figure}

We model the village effect as a spatial random effect. As in Example 4 from Section \ref{s:examples}, the linear predictor is
\begin{equation}
    \mbox{logit}\{P(Y_i=1|\eta_i)\} = \eta_i = \beta_0 + \bX_i\bbeta +u_{g_i}
\end{equation}
where $g_i\in\{1,\dots,L\}$ is the village of response $i$. We also assume that $\mbox{E}(u_i)=0$ and $\mbox{Var}(u_i)=\sigma^2_u$ for all $i$ and exponential spatial correlation $C_{ij}=\mbox{Cor}(u_i,u_j) = e^{-d_{ij}/\rho}$ where $d_{ij}$ is the distance between village $i$ and $j$ and $\rho>0$ is the spatial range parameter. Then the full prior specification for R2D2 is
\begin{multline}
       \beta_0\sim\mbox{Normal}(\mu_0,\tau_0^2),\
       \bbeta|\phi_1,W\sim\mbox{Normal}(0, \tfrac15\phi_1W\bI_5),\
       {\bf u}|\phi_2,W,\rho\sim\mbox{Normal}(0,\phi_2 W\bC),\\ \rho\sim\mbox{Uniform}(0, 2r), W\sim\mbox{GBP}(a^*, b^*, c^*, d^*),\ \bphi\sim\mbox{Dirichlet}(\xi_1,\xi_2)
\end{multline}
for hyper-parameters set to $\mu_0=0,\tau_0^2=3$, $\xi_1=\xi_2=1$ and $r$ is the maximum distance between pairs of villages. Note that $\sigma^2_u=\phi_2W$ in this model.  We find $\hat\beta_0=-0.59$ and $(a^*,b^*,c^*,d^*)$ are in Table \ref{tab:abcd} and the resulting prior distributions are plotted in Figure \ref{fig:gamb.rand.prior}. 

\begin{table}
    \centering
    \begin{tabular}{cc|cccc}
         $a$& $b$ & $a^*$&$b^*$&$c^*$&$d^*$  \\\hline
         1 & 4&1.15& 2.08& 0.91& 2.09\\
         0.5 & 0.5 &0.57 &0.29& 0.90& 1.54\\
         1 & 1&1.47& 0.65& 0.79& 1.67\\
         4 & 4&7.45& 2.72& 0.73& 1.63\\
         4 & 1&7.77& 0.71& 0.68 &1.45
    \end{tabular}
    \caption{Generalized Beta Prime approximation parameters  for {\tt Gambia} data with $\hat\beta_0=-0.59$.}
    \label{tab:abcd}
\end{table}

For PC prior, the full prior specification is
\begin{multline}
       \beta_0\sim\mbox{Normal}(\mu_0,\tau_0^2),\        \bbeta\sim\mbox{Normal}(0,\tau_1^2\bI_5),\
       {\bf u}|\sigma^2_u\sim\mbox{Normal}(0,\sigma^2_u\bC),\\ \rho\sim\mbox{Uniform}(0, 2r), \sigma_u\sim \mbox{Exp}(\lambda_0).
\end{multline}
where $\mu_0=0,\tau_0^2=3,\tau_1^2=100$ and $\lambda_0=-\log(0.01)/.968$. The vague prior has the same form as the PC prior except $\sigma^2_u\sim\mbox{InvGamma}(0.5,0.0005)$ \citep{winbugs}.  

We take 105,000 MCMC samples with the first 5,000 discarded as burn-in, where the analysis is performed in \texttt{JAGS}. The results are in Figure 
\ref{fig:gamb.spatial.post} and Table \ref{tab:gamb_post_s}. We also present trace plots in the Supplemental Materials to check convergence of the MCMC chain, as well as the effective sample size and computation time. 
We can see that the posterior distributions of $R_n^2$ are very similar across the different methods. The posterior of $W$, however, varies across the different R2D2 priors with the Beta(4,1) and Beta(4,4) having the greatest mean and Beta(0.5,0.5) and Beta(1,1) having the smallest mean. 
The posterior distributions of $W$ and $\sigma^2_u$ are almost identical for the R2D2 priors which means that the vast majority of the global variance mass is shifted on the random effect variance and away from the fixed effect variance. 
The posterior for $\sigma^2_u$ has the smallest mean for the PC prior, which follows from the fact that this prior shrinks the spatial variance toward zero. Lastly, the posterior of $\rho$ varies across the different priors with the PC prior also yielding the smallest posterior mean. 

\begin{figure}
    \centering
    \includegraphics[width=\textwidth]{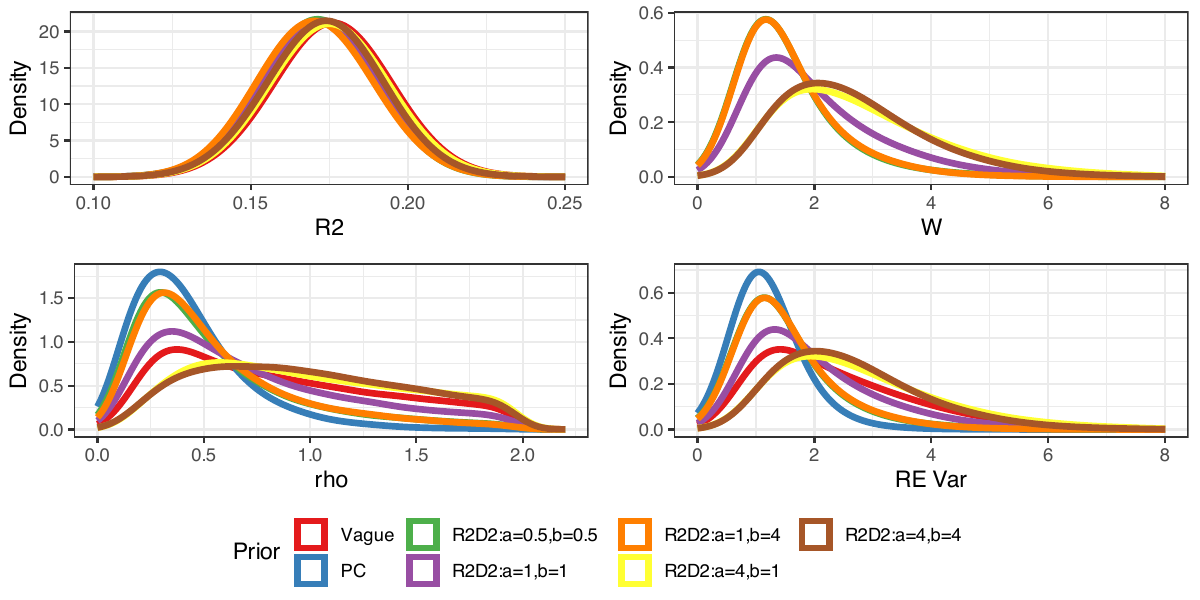}
    \caption{Posterior $R^2$, global variance and random effect variance and $\rho$ for vague (uninformative) prior distributions, PC and R2D2 for {\tt Gambia} data with village spatial random effect.}
    \label{fig:gamb.spatial.post}
\end{figure}

\begin{table}
    \centering
 \begin{tabular}{c|cc|cc|cc|cc}
    \multicolumn{1}{c|}{} &
    \multicolumn{2}{c|}{$R^2$}& 
    \multicolumn{2}{c|}{$W$}&
    \multicolumn{2}{c|}{$\sigma^2_u$}&
    \multicolumn{2}{c}{$\rho$}\\
         Method & Mean & St. Dev & Mean & St. Dev & Mean & St. Dev  & Mean & St. Dev \\\hline
        Vague & 0.176 &0.016 &$-$ &$-$ &2.389 &1.397 &0.851 &0.500\\
        PC &0.173 &0.016 &$-$ & $-$ &1.210 &0.515 &0.427 &0.270 \\
        $R^2\sim\mbox{Beta}(\tfrac12,\tfrac12)$ & 0.171 &0.015 &1.538 &0.827 &1.511 &0.824 &0.549 &0.388\\
        $R^2\sim\mbox{Beta}(1,1)$&0.173 &0.016 &2.023 &1.148 &1.992 &1.143 &0.722 &0.465  \\
        $R^2\sim\mbox{Beta}(1,4)$ & 0.171 &0.016 &1.540 &0.796 &1.513 &0.792 &0.556 &0.382 \\
        $R^2\sim\mbox{Beta}(4,1)$ &0.175 &0.016 &2.813 &1.412 &2.780 &1.409 &0.977 &0.488\\ 
        $R^2\sim\mbox{Beta}(4,4)$ &0.174 &0.016 &2.675 &1.237 &2.637 &1.227 &0.981 &0.479
    \end{tabular}
    \caption{Posterior mean and standard deviation for $R_n^2$, global variance ($W$), random effect variance ($\sigma^2_u$) and spatial range ($\rho$) for each method for {\tt Gambia} data considering spatial random effect.}
    \label{tab:gamb_post_s}
\end{table}

\subsection{Genomics data}

For our second data analysis, we apply the proposed method to high-dimensional genomics data from human breast tumors \citep{perou2000molecular} in the \texttt{mixOmics} R package \citep{rohart2017mixomics}. This data was pre-processed in \cite{perez2003prediction} such that the response, $Y_i$, is 1 if tumor specimen $i$ was analyzed before chemotherapy treatment, and 0 if it was analyzed after chemotherapy, for sample $i\in\{1,\dots,n\}$ where $n=47$. There are $p=1000$ gene expressions as predictor variables, so $p\gg n$, and we imputed the appropriate column mean of $\bX$ for any missing values. We consider the R2D2 prior with $(a,b)\in\{(1, 5),(1,10), (1,20), (1,30)\}$, and compare with the Horseshoe prior. We choose these hyper-parameter combinations because we hypothesize that prior distributions with large mass near $R^2=0$ are useful in high-dimensional contexts.

Since our focus of this analysis is prediction, we randomly split the data into train and test sets, where 75\% of the data is used for training, and 25\% for testing. The model is fit on the training data, and we compute the Brier Score (BS) and binary cross entropy (BCE) loss on the hold-out test data. If $Y_i$ is the true response, and $\hat Y_i$ is the predicted response, then
$
    \mbox{BS}=\frac1n\sum_{i=1}^n (\hat Y_i-Y_i)^2
$
and
$
    \mbox{BCE}=-\frac1n\sum_{i=1}^n \{Y_i\log\hat Y_i+(1-Y_i)\log(1-\hat Y_i)\}.
$
We compute $\hat Y_i=[1+\exp\{-(\hat\beta_0+\bX'_i\hat\bbeta)\}]^{-1}$ where $\hat\beta_0$ and $\hat\bbeta$ are the posterior means of $\beta_0$ and $\bbeta$, respectively, and $\bX_i'$ are the predictor variables for the $i$th test data. The splitting procedure is repeated 50 times and we find the average of each metric. A good method will have small BS and BCE. The analysis was performed in \texttt{Stan} where we computed 10,000 MCMC samples with an additional 1000 for burn-in.

The results are in Table \ref{tab:genom}. In the Supplemental Materials, we also report the computation time, and average effective sample size for $\bbeta$ and the global variance parameter. We can see that the R2D2 prior with $(a,b)=(1,20)$ and Horseshoe perform the best, with smallest BCE and BS, respectively. Indeed, there is a clear trend that the R2D2 priors with more mass near $R^2=0$ yield better predictions. However, there appears to be diminishing returns beyond $R^2\sim\mbox{Beta}(1,10)$.

\begin{table}[]
    \centering
    \begin{tabular}{l|cc}
        Method & BS & BCE \\\hline
        Horseshoe & {\bf 0.23} (0.01) & 0.79 (0.07)\\
        $R^2\sim\mbox{Beta}(1,5)$ & 0.29 (0.01) & 0.89 (0.06) \\
        $R^2\sim\mbox{Beta}(1,10)$& 0.26 (0.00) & 0.71 (0.01) \\
        $R^2\sim\mbox{Beta}(1,20)$& 0.26 (0.00) & {\bf 0.70} (0.01)\\
        $R^2\sim\mbox{Beta}(1,30)$& 0.26 (0.00) & 0.71 (0.01)
    \end{tabular}
    \caption{Prediction results for genomics data with standard error in parentheses. BS is Brier Score, and BCE is the binary cross entropy loss. Best results are labeled in {\bf bold}. }
    \label{tab:genom}
\end{table}

\section{Discussion}\label{s:discussion}

In this work, we proposed a novel method for choosing informative prior distributions in the generalized linear mixed model setting. The proposed prior is flexible and interpretable in terms of overall model fit as measured by a Bayesian $R^2$. There are many cases where the prior $R^2$ can be induced exactly as well as general approximation strategies when an exact form is not possible. The main approach that we suggest is approximating the global variance prior with a generalized beta prime distribution because of its flexibility and ability to be implemented in standard software. Combined with an initial estimate of the intercept via a method of moments estimator and the GBP approximation in the \texttt{r2d2glmm} package, we provide a simple and intuitive method for setting prior distributions in GLMMs.

If there is domain knowledge available on how well the model is expected to fit the data, then this could be used to inform prior choice for $R^2$.
In the absence of any prior information, we suggest $R^2\sim\mbox{Beta}(1,1)$ as a reasonable default choice. This prior expresses a full range of model fits, but should not be confused with a flat prior for $\bbeta$. Choosing $R^2\sim\mbox{Beta}(1,b)$ for large $b$, or another prior with large mass near 0, is also a good choice, especially when working in a high-dimensional setting. Indeed, substantial prior mass near $R^2=0$ prevents the model from overfitting by shrinking towards the ``base model'' of $\bbeta={\bf 0}_p$. 

The proposed approach naturally fits within the global-local shrinkage prior framework where $W$ controls the global shrinkage, and $\phi_j$ the local shrinkage. Our approach does differ from, e.g., \cite{hamura2022global}, who studies shrinkage priors for count data. Using the notation of our paper, \cite{hamura2022global} let $Y_i|\lambda_i\sim\mbox{Poisson}(\lambda_i e^{\eta_i})$ where $\lambda_i\sim\mbox{Gamma}(\alpha,\beta/v_i)$, $v_i\sim\pi(\cdot)$, and $\eta_i$ is some offset which may depend on covariates, i.e., $\eta_i=\beta_0+\bX_i\bbeta$. This paper is primarily interested in the prior distribution for the local component ($v_i$) and estimating the rate parameter ($\lambda_i$). The R2D2 prior, on the other hand, is derived for the global component ($W$), and is focused on estimating the regression components ($\eta_i$).

A limitation of the proposed method is that the hierarchical framework only allows for random intercepts and not, for example, random slopes. Additionally, the finite mean and variance requirement precludes applications to some models, e.g., extreme value analysis \citep{coles2001introduction}. We have also not proven concentration or shrinkage properties which is an avenue for future work. We could also extend the method to allow for other survival analysis settings beyond the uncensored Weibull model and models that are not GLMMs such as Bayesian deep learning. Finally, at present, the method cannot easily be written in \texttt{INLA} \citep{lindgren2015bayesian}, so we consider this an important next step.

\section*{Supplemental Materials}

In the Supplemental Materials, we provide: further discussion and comparison of the different coefficient of determination definitions, additional models and derivations for priors with closed-form expressions, a table showing different values of $(a^*,b^*,c^*,d^*)$ for the GBP approximation, a simulation study, and additional results for the real-data analysis.
\par
\section*{Acknowledgements}
The authors thank Brandon Feng for help with the derivation of the Weibull model. We also thank the National Institutes of Health (R01ES031651-01) and King Abdullah University of Science and Technology (3800.2) for financial support.

\bibliographystyle{chicago}      
\bibliography{refs}   

\begin{center}
{\Large Supplemental Materials: The R2D2 prior for generalized linear mixed models}
\end{center}

In this Supplemental Materials, we provide: further discussion and comparison of the different coefficient of determination definitions, additional models and derivations for priors with closed-form expressions, a table showing different values of $(a^*,b^*,c^*,d^*)$ for the GBP approximation, a simulation study, and additional results for the real-data analysis.

\section{Discussion on $R^2$}
\subsection{Comparison of sample and population $R^2$}
In this sub-section, we compare the sample and population definition of $R^2$ (in (3.1) and (3.2), respectively) under the location-scale model in Section 3.1.  In this model, 
$\boldsymbol{\eta}|\beta_0,W\sim\mbox{Normal}(\bbeta_0,W\Sigma)$ for correlation matrix $\Sigma$ and $\mu(\eta_i)=\eta_i$ and $\sigma^2(\eta_i)=\sigma^2$ for $i=\{1,\dots,n\}$.  Since the mean operator simplifies to $M\{\sigma^2(\eta_1),...,\sigma^2(\eta_n)\} = \sigma^2$, $R^2_n$ converges in probability to $R^2$ if and only if $V\{\eta_1,...,\eta_n\}$ converges in probability to $W$. Since our prior distributions are conditioned on $\beta_0$ and $W$, we take the variance operator to be 
$ v_n=V\{\eta_1,...,\eta_n\} = (\boldsymbol{\eta}-\beta_0\boldsymbol{1}_n)^T(\boldsymbol{\eta}-\beta_0\boldsymbol{1}_n)/n$.  From the properties of quadratic forms, we have $\mbox{E}(v_n)=W$ and $\mbox{V}(v_n)=2W^2\mbox{tr}(\Sigma\Sigma)/n^2$.  Therefore, if $tr(\Sigma\Sigma)=o(n^2)$ then $R_n^2\rightarrow W/(W+\sigma^2)=R^2$.

The key term is $\mbox{tr}(\Sigma\Sigma)$, which simplifies in the linear model $\boldsymbol{\eta}=\beta_0+\bX\bbeta$.  Then $v_n=\bbeta^T\bX^T\bX\bbeta/n$ with $\bX^T\bX=(n-1)\bR_X$ and $\bR_X$ being the $p\times p$ sample covariance matrix of $\bX$, and $\Sigma = \bX\bP\bX^T$ for diagonal matrix $\bP$ with diagonal elements $\{\phi_1,...,\phi_p\}$.  If we further assume that $\bR_x = \boldsymbol{I}_p$ and $\phi_j=1/p$, then $\mbox{tr}(\Sigma\Sigma) = (n-1)^2/p$.  Thus, $\mbox{tr}(\Sigma\Sigma)=o(n^2)$ if and only if $p$ diverges with $n$.  In this special case, the number of free parameters increases. The intuition is that since we are conditioning on $\bX$, the random quantity is $\bbeta$, and the sample variance converges to the true variance $W$ only when the number of random variables in $\bbeta$ increases. Of course, this is only one special case, but even in this simple case, it is informative to see the dependence on diverging $p$. Meanwhile, the condition $tr(\Sigma\Sigma)=o(n^2)$ provides further insight to study the finite sample and population versions.

We note that in practice we use the sample mean $\bar{\eta} = \sum_{i=1}^n\eta_i/n$ in the variance operator, $V(\eta_1,...,\eta_n) = (\boldsymbol{\eta}-\bar{\eta}\boldsymbol{1}_n)^T(\boldsymbol{\eta}-\bar{\eta}\boldsymbol{1}_n)/(n-1)$. Using this definition and the location-scale model above, it can be shown that $R^2_n$ converges in probability to $R^2$ if and only if ${\bf1}_n'\Sigma{\bf1}_n=o(n^2)$ and $\mbox{tr}(\bA_n\bA_n)=o(n^2)$, where $\bA_n=(\bI_n-\bP_n)\Sigma$ and $\bP_n=\frac1n{\bf 1}_n{\bf 1}_n'$. For example, in a one-way random effects model, these conditions are met if and only if the number of levels of the random effect diverges.

Finally, in the linear mixed model, we allow for random intercepts. Since these random effects have mean 0, then, on average, we still have $\bar{\eta}=\beta_0$ and thus convergence results are equivalent to those for the linear model above requiring $tr(\Sigma\Sigma)=o(n^2)$. In this subsection, we have only considered a few special cases so we suggest being aware of the possible discrepancy between the population and sample $R^2$ definitions.

\subsection{Comparison with other measures of $R^2$}
We also provide a brief comparison of the sample and population versions of $R^2$ with some other measures of model fit for GLMMs. First, recall that
$$
    R^2=\frac{\mbox{Var}\{\mu(\eta)\}}{\mbox{Var}\{\mu(\eta)\}+\mbox{E}\{\sigma^2(\eta)\}}
$$
where $\mu(x)=\{1+\exp(-x)\}^{-1}$ and $\sigma^2(x)=\mu(x)\{1-\mu(x)\}$. Additionally,
$$
    R^2_n
    =\frac{\frac{1}{n-1}\sum_{i=1}^n \{(\mu(\eta_i)-\bar\mu\}^2}{\frac{1}{n-1}\sum_{i=1}^n \{(\mu(\eta_i)-\bar\mu\}^2 + \frac1n\sum_{i=1}^n \sigma^2(\eta_i)}
$$
where $\bar\mu=\frac1n\sum_{i=1}^n\mu(\eta_i)$.

Now, we compare these with two popular measures for logistic regression: $R^2_{CS}$ and $R^2_M$. $R^2_{CS}$ was proposed by \cite{cox1989analysis} and is defined as
\begin{equation}\label{eq:cs}
    R^2_{CS}
    =1-\left(\frac{L_0}{L_M}\right)^{2/n}
\end{equation}
where $L_0$ and $L_M$ are the likelihoods of the intercept-only and full model, respectively. For example, under the logistic regression model, \eqref{eq:cs} becomes
$$
    R^2_{CS}
    =1- \prod_{i=1}^n \left\{\frac{\mu(\beta_0)}{\mu(\eta_i)}\right\}^{y_i}\left\{\frac{1-\mu(\beta_0)}{1-\mu(\eta_i)}\right\}^{1-y_i}.
$$
\cite{mcfadden1973conditional} proposed $R^2_M$ defined as 
\begin{equation}\label{eq:m}
    R^2_M
    =1-\frac{\log L_M}{\log L_0}.
\end{equation}
Using the logistic regression likelihood, \eqref{eq:m} is
$$
    R^2_M= 1-\frac{\sum_{i=1}^n y_i\log\mu(\eta_i)+(1-y_i)\log\{1-\mu(\eta_i)\}}{\sum_{i=1}^n y_i\log\mu(\beta_0)+(1-y_i)\log\{1-\mu(\beta_0)\}}.
$$

There are some similarities and differences between the different measures. First, each metric provides a measure of the variation in the response that is explained by the model. In particular, a value of 0 corresponds to the base (intercept-only) model for each metric. There are, however, some key differences. Most importantly, $R^2_{CS}$ and $R^2_M$ depend on the model's likelihood, and therefore also depend on the response $Y_i$. This is not a problem for an {\it a posteriori} measure of model fit, but for setting prior distributions this is problematic, as typically the response is unavailable when choosing prior distributions. For these reasons, along with it being much simpler to derive the prior distribution for $R^2$, we leverage this definition in this work.

\section{Models with exact prior forms}
This is a continuation of Section 3.1 where we derive the exact prior distribution for several models.

\paragraph{Poisson regression with offsets:}
Poisson regression models often include a fixed offset term $N_i$, e.g., if $i$ is a spatial region then $N_i$ may be taken as the population of region $i$. The model is $Y_i|\eta_i\sim \mbox{Poisson}(e^{\eta_i})$ where $\eta_i = \log(N_i)+\beta_0 + \bX_i\bbeta+\sum_{k=1}^qZ_{ik}\bu_k$. As with the other covariates, we standardize the log offset terms so that $\sum_{i=1}^n\log(N_i)=0$ and $\Var\{\log(N_i)\}=\sigma^2_N$ and treat the offset as a random variable independent of each of the other terms in the model. Thus, $\eta|\beta_0,W\sim\mbox{Normal}(\beta_0,W+\sigma^2_N)$ so
\begin{equation}
    R^2
    =\frac{\theta e^{W}-1}{\theta e^{W}-1+\theta^{-1/2}e^{-\beta_0-\frac12W}}.
\end{equation}
where $\theta=e^{\sigma^2_N}$. Because variability in the offset terms remains even if $W=0$, the lower bound of $R^2$ is
\begin{equation}
    R^2_{min}
    =\frac{\theta-1}{\theta-1+\theta^{-1/2}e^{-\beta_0}} >0.
\end{equation}
In this case, we use the four-parameter beta prior $\tilde R^2\sim\mbox{Beta}(a,b,R^2_{min},1)$ conditioned on $\beta_0$ and $\theta$ that induces the prior for $W$ with density
\begin{multline}
        \pi(w|\beta_0,\theta; a,b)
    =\frac{\theta^{a/2} e^{a\beta_0}\{1+e^{\beta_0}(\theta-1)\sqrt{\theta}\}^b}{2B(a,b)}\\
    \times \frac{\{1-\theta+e^{w/2}(\theta e^{w}-1)\}^{a-1} (3\theta e^{3w/2}-e^{w/2})}{ \{1+\sqrt{\theta}e^{\beta_0+w/2}(\theta e^{w}-1)\}^{a+b}},\ w\geq0.
\end{multline}

\paragraph{Negative Binomial regression:} The Negative Binomial (NB) distribution generalizes the Poisson distribution and allows for overdispersion. Let $Y|\eta,\theta\sim\mbox{NB}(e^\eta,\theta)$, parameterized so that $\mu(\eta) = e^{\eta}$ and $\sigma^2(\eta) = \theta e^{\eta}$ for overdispersion parameter $\theta>1$.
Similar to the Poisson example, 
\begin{equation}
    R^2
    =\frac{e^{W}-1}{e^{W}-1+\theta e^{-\beta_0-\frac12W}}.
\end{equation}
$R^2\sim\mbox{Beta}(a,b)$ induces (see Supplementary Materials) the prior for $W$, conditioned on $\theta$, with density
\begin{equation}\label{eq:OD}
    \pi(w|\beta_0, \theta; a,b)
    =\frac{\theta^b}{2B(a,b)} \frac{(e^w-1)^{a-1}e^{-b(\beta_0+w/2)}(3e^w-1)}{(e^w-1+\theta e^{-\beta_0-w/2})^{a+b}},\ w\geq 0.
\end{equation}
The shape of the prior is very similar to that of the Poisson case, except that is has a greater probability of a larger value. The value of the prior at 0 is $\infty$ if $a<1$, $be^{\beta_0}/\theta$ if $a=1$ and $0$ if $a>1$.

\paragraph{Zero-inflated Poisson regression:} Another generalization of the Poisson model is the zero-inflated Poisson (ZIP) model. In the ZIP model, $Y|\eta$ is zero with probability $\pi(\eta)$ and Poisson with mean $\lambda(\eta)$ with probability $1-\pi(\eta)$.  Then $\mu(\eta)=\{1-\pi(\eta)\}\lambda(\eta)$ and $\sigma^2(\eta)=\{1-\pi(\eta)\}\lambda(\eta)\{1+\pi(\eta)\lambda(\eta)\}$. A closed form solution for the R2D2 prior exists for the special case with $\pi(\eta) = \theta$ for all $\eta$ and $\lambda(\eta)=e^\eta$.  Then
\begin{equation}
   R^2 = \frac{(1-\theta)(e^W-1)}{(1-\theta)(e^W-1)
     + e^{-\beta_0-W/2} + \theta e^{W}}.
\end{equation}
In this case, $R^2$ is bounded above by $R^2_{max}=1-\theta$ so $\tilde R^2\sim\mbox{Beta}(a,b,0,1-\theta)$ induces the prior for $W$ with density:
\begin{multline}
    \pi(w|\beta_0,\theta;a,b) 
    = \frac{(e^w-1)^{a-1} e^{-b(\beta_0+w/2)}(1+\theta e^{\beta_0+3w/2})^b}
    {2B(a,b)(e^w-1+e^{-\beta_0-w/2}+\theta)^{a+b}} \\
    \times \frac{(3e^w-1+2\theta e^{\beta_0+3w/2})}
    {(1+\theta e^{\beta_0+3w/2})},\ w\geq0.
\end{multline}
The value of the prior at 0 is $\infty$ if $a<1$, $be^{-b\beta_0}(1+\theta e^{\beta_0})^b/(\theta+e^{-\beta_0})^{1+b}$ if $a=1$ and 0 if $a>1$.

\paragraph{Weibull model:} Consider the Weibull model (without censoring) $Y|\eta,\theta\sim \mbox{Weibull}(e^\eta,\theta)$ such that $\mu(\eta)=e^{\eta}\Gamma\left(1+\frac1\theta\right)$ and $\sigma^2(\eta)=e^{2\eta}\left\{\Gamma\left(1+\frac2\theta\right)-\Gamma^2\left(1+\frac1\theta\right)\right\}$ for shape parameter $\theta>0$. Then
\begin{equation}
    R^2
    =\frac{e^W-1}{\frac{\Gamma(1+\frac2\theta)}{\Gamma^2(1+\frac1\theta)}e^W-1}.
\end{equation}
Interestingly, this does not depend on $\beta_0$. $R^2$ is bounded above by $R^2_{max}=\Gamma^2(1+\frac1\theta)/\Gamma(1+\frac2\theta):=r^{-1}(\theta)$ so $\tilde R^2\sim \mbox{Beta}(a,b,0,r^{-1}(\theta))$ induces a prior for $W$ with density:
\begin{equation}
    \pi(w|\theta;a,b)
    =\frac{\{r(\theta)-1\}^b}{B(a,b)}\frac{e^{bw}(e^w-1)^{a-1}}{\{r(\theta)e^w-1\}^{a+b}},\ w\geq0.
\end{equation}
The value of the prior at 0 is $\infty$ if $a<1$, $b/\{r(\theta)-1\}$ if $a=1$ and 0 if $a>1$.

\section{Derivations} 
\noindent{\bf Derivation of (2.2)}: There are two ways to think of $\eta_i$ being normally distributed. First, we consider the case where $\bX_i$ is treated as random with mean $\boldsymbol\mu$ and variance $\Sigma_{{\bf X}}$. For theoretical convenience, we assume that $\boldsymbol\mu={\bf 0}_p$ and $diag(\Sigma_{{\bf X}})={\bf 1}_p$. In practice, $\bX$ can be empirically standardized such that each column has mean zero and variance one. Then, for moderate $p$, $\bX_i\bbeta$ will be approximately normally distributed by the Central Limit Theorem. Thus, $\eta_i$ is a linear combination of normal random variables so it too will be distributed normally and we simply must find the mean and variance. The mean is
\begin{align}
    \E(\eta_i|\beta_0,W,\boldsymbol\phi)
    &=\beta_0 + \E(\bX_i\bbeta|\beta_0,W,\boldsymbol\phi)
    + \E(\sum_{k=1}^q u_{kg_{ik}}|\beta_0,W,\boldsymbol\phi)\\
    &=\beta_0 + \E_{\bX_i}\{\bX_i\E_{\bbeta}(\bbeta|\bX_i,\beta_0,W,\boldsymbol\phi)\} + \sum_{k=1}^q  \E(u_{kg_{ik}}|\beta_0,W,\boldsymbol\phi)=\beta_0.\nonumber
\end{align}
The variance is $\Var(\eta_i|\beta_0,W,\boldsymbol\phi)
    =\Var(\bX_i\bbeta|\beta_0,W,\boldsymbol\phi) +\Var(\sum_{k=1}^q u_{kg_{ik}}|\beta_0,W,\boldsymbol\phi)$. The first term is
\begin{align}
    \Var(\bX_i\bbeta|\beta_0,W,\boldsymbol\phi)
    &=\E_{\bX_i}\{\Var_{\bbeta}(\bX_i\bbeta|\bX_i,\beta_0,W,\boldsymbol\phi)\} + \Var_{\bX_i}\{\E_{\bbeta}(\bX_i\bbeta|\bX_i,\beta_0,W)\}\\
    &=\E_{\bX_i}\{\bX_i [W diag(\phi_1,\dots,\phi_p)]\bX_i^T|\beta_0,W,\boldsymbol\phi\} + \Var_{\bX_i}(0|\beta_0,W,\boldsymbol\phi)\nonumber\\
    &=\E_{\bX_i}(\mbox{tr}\{\bX_i [W diag(\phi_1,\dots,\phi_p)]\bX_i^T\}|\beta_0,W,\boldsymbol\phi)\nonumber\\
    &=W\mbox{tr}\{diag(\phi_1,\dots,\phi_p)\E(\bX_i^T\bX_i|\beta_0,W,\boldsymbol\phi)\}\nonumber\\
   &=W\mbox{tr}\{diag(\phi_1,\dots,\phi_p)\Sigma_{\bX}\}\\
   &=W\sum_{j=1}^p \phi_j.\nonumber
\end{align}
Similarly, the second term is 
\begin{align}
    \Var(\sum_{k=1}^q u_{kg_{ik}}|\beta_0,W,\boldsymbol\phi)
    =\sum_{k=1}^q \Var(u_{kg_{ik}}|\beta_0,W,\boldsymbol\phi)
    =W\sum_{k=1}^q\phi_{p+k}
\end{align}
Combining these two terms gives $\Var(\eta_i|\beta_0,W,\boldsymbol\phi)
    = W\sum_{j=1}^p\phi_j +W\sum_{k=1}^q \phi_{p+k}=W
$.

On the other hand, we can treat $\bX_i$ as fixed where each column is again standardized to have mean zero and variance one. Then $\eta_i$ is a linear combination of normal random variables so it too will be normally distributed with the following mean and variance:
$$
    \E(\eta_i|\beta_0,W,\boldsymbol\phi)
    =\beta_0 + \bX_i\E(\bbeta|\beta_0,W,\boldsymbol\phi)
    + \E(\sum_{k=1}^q u_{kg_{ik}}|\beta_0,W,\boldsymbol\phi)
    =\beta_0,
$$
and
\begin{align*}
    \Var(\eta_i|\beta_0,W,\boldsymbol\phi)
    &=\bX_i\mbox{Var}(\bbeta|\beta_0,W,\boldsymbol\phi)\bX_i^T + \sum_{k=1}^q \mbox{Var}(u_{kg_{ik}}|\beta_0,W,\boldsymbol\phi) \\
    &=W\sum_{j=1}^p \phi_j x_{ij}^2 + W\sum_{k=1}^q \phi_{p+k}.
\end{align*}
Now, notice that this variance is different for each $\eta_i$ since it depends on $x_{ij}$. Therefore, we can consider the {\it average} variance (over all observations) and we find:
$$
    \frac1n\sum_{i=1}^nW\sum_{j=1}^p \phi_j x_{ij}^2 + \frac1n\sum_{i=1}^nW\sum_{k=1}^q \phi_{p+k}
    =W\sum_{j=1}^p\phi_j\frac1n\sum_{i=1}^n x_{ij}^2 + W\sum_{k=1}^q \phi_{p+k}
    \approx W\sum_{j=1}^{p+q}\phi_j
    =W
$$
because $\bX$ is standardized such that $\sum_{i=1}^n x_{ij}^2=n-1$ for all $j$. In this way, the average distribution of $\eta_i|\beta_0,W,\boldsymbol\phi\sim\mbox{Normal}(\beta_0,W)$.\\

\noindent{\bf Derivation of (3.8)}: We have
$$    f_R(r)
    =\frac{1}{B(a,b)} r^{a-1}(1-r)^{b-1},\ 0\leq r\leq 1
$$
Now, 
$$    R^2
    =g^{-1}(W)
    =\frac{e^W-1}{e^W-1+e^{-\beta_0-W/2}}
$$
So,
$$    \frac{d}{dw}g^{-1}(w)
    =\frac{e^{-\beta_0-w/2}(3e^w-1)}{2(e^W-1+e^{-\beta_0-W/2})^2}
$$
Thus,
\begin{align}
    f_W(w)
    &=\frac1{B(a,b)}\left(\frac{e^w-1}{e^w-1+e^{-\beta_0-w/2}}\right)^{a-1}\left(\frac{e^{-\beta_0-w/2}}{e^w-1+e^{-\beta_0-w/2}}\right)^{b-1}\cdot\frac{e^{-\beta_0-w/2}(3e^w-1)}{2(e^W-1+e^{-\beta_0-W/2})^2}\nonumber\\
    &=\frac{1}{B(a,b)} \frac{(e^w-1)^{a-1} e^{-b(\beta_0+w/2)}(3e^w-1)}{2(e^w-1+e^{-\beta_0-w/2})^{a+b}},\ w\geq0
\end{align}

\noindent{\bf Derivation of Supplementary Materials (4)}. Conditioning on $\theta$,
$$    R^2
    =g^{-1}(W)
    =\frac{e^W-1}{(1+\theta)e^W-1 + e^{-\beta_0-\frac12W}},
$$so
$$    \frac{d}{dw}g^{-1}(w)
    =\frac{e^{-\beta_0 - w/2} (2 \theta e^{\beta_0 + 3 w/2} + 3 e^w - 1)}{2 ( (\theta + 1)(e^w - 1) + e^{-\beta_0-w/2})^2}.
$$Thus,
\begin{align}
    f_W(w|\theta)
    &=\frac1{B(a,b)}\left(\frac{e^w-1}{(1+\theta)e^w-1 + e^{-\beta_0-w/2}}\right)^{a-1}\left(\frac{\theta e^w + e^{-\beta_0-w/2}}{(1+\theta)e^w-1 + e^{-\beta_0-w/2}}\right)^{b-1}\nonumber\\
    &\times\frac{e^{-\beta_0 - w/2} (2 \theta e^{\beta_0 + 3 w/2} + 3 e^w - 1)}{2 ( (\theta + 1)(e^w - 1) + e^{-\beta_0-w/2})^2}\\
    &= \frac1{2B(a,b)} \frac{e^{-\beta_0 - w/2}(e^w-1)^{a-1}(\theta e^w+e^{-\beta_0-w/2})^{b-1} (2 \theta e^{\beta_0 + 3 w/2} + 3 e^w - 1)}{\{(1+\theta)e^w-1+e^{-\beta_0-w/2}\}^{a+b}},\ w\geq0.\nonumber
\end{align}
All other distributions from this section are found similarly.

\section{Parameter values for generalized beta prime distribution}

In Table \ref{tab:GBP}, we present the GBP approximations for a selection of $(a,b)$ combinations for the Poisson, logistic and negative binomial (with overdispersion $\theta=2)$ models. In most cases, the best fitting $a^*$ and $b^*$ values are not close to $(a,b)$ which demonstrates the need for this approximation. Also notice that for Poisson, $c>1$ for many scenarios which means that it will have lighter tails than a Beta Prime, whereas for logistic, often times $c<1$ so these will have heavier tails than a Beta Prime.

\begin{table}
    \centering
    \begin{tabular}{ccc|cccc|cccc|cccc}
    \multicolumn{3}{c|}{Prior} &
    \multicolumn{4}{c|}{Poisson}& \multicolumn{4}{c|}{Logistic}&
    \multicolumn{4}{c}{Negative Binomial}\\
         $\beta_0$ & $a$ & $b$ & $a^*$ & $b^*$ & $c^*$ & $d^*$ & $a^*$ & $b^*$ & $c^*$ & $d^*$ & $a^*$ & $b^*$ & $c^*$ & $d^*$ \\\hline
         & $\frac12$ & $\frac12$ &
        0.19 & 0.77 & 4.22 & 3.17 & 0.48 & 0.22 & 1.23 & 1.78 & 0.21 & 0.74 & 4.78 & 3.49\\
         & 1 & 1 & 
          0.42 & 1.50 & 3.75 & 2.56 & 1.45 & 0.51 & 0.99 & 1.74 & 0.44 & 1.46 & 4.31 & 2.93\\
        $-2$ & 1 & 4 &
            0.36 & 4.29 & 3.32 & 1.98 & 0.99 & 1.72 & 1.19 & 2.53 & .36 & 4.98 & 3.95 & 2.51\\
         & 4 & 1 &
          2.81 & 2.61 & 2.84 & 2.43 & 8.21 & 0.65 & 0.74 & 1.49 & 2.50 & 2.04 & 3.65 & 2.76\\
         & 4 & 4 &
            2.00 & 6.38 & 3.14 & 2.25 & 8.15 & 2.18 & 0.88 & 1.57 & 3.50 & 6.99 & 2.95 & 2.45    \\\hline
         & $\frac12$ & $\frac12$ &
           0.23 & 0.96 & 2.31 & 2.03 & 0.72 & 0.39 & 0.85 & 1.31 & 0.20 & 0.87 & 2.98 & 2.47\\
         & 1 & 1 & 
            0.50 & 1.83 & 2.00 & 1.45 & 1.47 & 0.67 & 0.77 & 1.68 & 0.44 & 1.67 & 2.60 & 1.84\\
        0 & 1 & 4 &
            0.63 & 5.49 & 1.52 & 0.95 & 1.17 & 2.12 & 0.89 & 2.03 & 0.50 & 4.85 & 2.00 & 1.27\\
         & 4 & 1 &
            2.08 & 2.68 & 1.92 & 1.53 & 7.72 & 0.72 & 0.68 & 1.44 & 2.24 & 2.65 & 2.24 & 1.85\\
         & 4 & 4 &
            2.10 & 6.65 & 1.83 & 1.12 & 7.37 & 2.79 & 0.72 & 1.65 & 1.83 & 6.65 & 2.28 & 1.57    \\\hline
         & $\frac12$ & $\frac12$ &
            0.49 & 1.38 & 0.93 & 0.70 & 0.48 & 0.22 & 1.23 & 1.78 & 0.37 & 1.19 & 1.26 & 1.11\\
         & 1 & 1 & 
            0.99 & 2.33 & 0.94 & 0.38 & 1.45 & 0.51 & 0.99 & 1.74 & 0.80 & 2.27 & 1.16 & 0.71\\
        2 & 1 & 4 &
            1.14 & 1.98 & 0.89 & 0.44 & 0.99 & 1.72 & 1.19 & 2.53 & 0.96 & 8.19 & 1.03 & 0.55\\
         & 4 & 1 &
            2.38 & 2.77 & 1.11 & 0.53 & 8.21 & 0.65 & 0.74 & 1.49 & 2.16 & 2.78 & 1.35 & 0.87\\
         & 4 & 4 &
           3.66 & 9.86 & 0.97 & 0.36 & 8.15 & 2.18 & 0.88 & 1.57 & 2.92 & 6.44 & 1.24 & 0.43
    \end{tabular}
    \caption{Parameter values for Generalized Beta Prime distribution in order to approximately induce $R^2\sim\mbox{Beta}(a,b)$. Negative binomial takes $\theta=2$.}
    \label{tab:GBP}
\end{table}

\section{Simulation study}\label{s:sim}

Here we apply the methods described in Section 3 to simulated data. The objectives are to compare the proposed method with other methods, as well as understand how the proposed method performs under different combinations of $(a,b)$. The different combinations of $(a,b)$ that we compare are $(1,1), (1,4)$ and $(4,1)$ using the GBP approximation of Section 3.2.3.

We consider simulations for linear regression with random effects as \cite{zhang2022bayesian} already considered the case of fixed effects and sparsity. For the generalized linear models, we consider two cases: Poisson regression with mixed effects and high-dimensional Logistic regression with fixed effects. Throughout these experiments, we consider a range of true $R^2$ values from 0.35 to 0.66.

We compare the proposed method to two leading methods. For mixed effects cases, we consider the penalized complexity (PC) prior of \cite{simpson2017penalising} and for the fixed effects case we consider the horseshoe prior of \cite{carvalho2010}. We also compare with a simple vague prior. Details of the priors are given below.

We use several metrics of comparison. First, we measure the bias and mean squared error (MSE) of the observed $R^2$. We compute $\hat R^2_n$ using (3.3) and the true value by plugging in the true values of fixed and random effects into the definition in (3.3). We also compute the difference between the true $\bbeta$ and estimated $\hat\bbeta$,
$
    ||\hat\bbeta-\bbeta||_2
    =\sum_{j=1}^p (\hat\beta_j-\beta_j)^2/p.
$
For the random effects scenarios, we compute the MSE of the estimated random effect variances. Lastly, we measure the performance of the method as computed by prediction error on hold-out test data, $\tilde Y$ and fitted values $\hat Y$, both of size $N=1000$. In the Gaussian case, we compute the MSE as $\frac1N\sum_{i=1}^N(\tilde Y_i-\hat Y_i)^2$. In the Poisson case, we compute the log-score as
$
    \frac1N\sum_{i=1}^N \log\{f(\tilde Y_i;\lambda=\hat Y_i)\}
$
where $f(\cdot|\lambda)$ is the probability mass function for a Poisson$(\lambda)$ random variable. For the Logistic case, we compute the area under the receiver operator curve (AUC). In each setting we simulate 200 data sets and take the average and standard error of these metrics. For all methods we use {\tt JAGS} \citep{plummer2016} for posterior computation with 10,000 MCMC samples where the first 5,000 are discarded as burn-in.

\subsection{Gaussian regression with random effects}
Let $\beta_0=1$ and consider two-way random effects without interaction with $u_{1i}\sim\mbox{Normal}(0,\sigma^2_{u_1})$ for $i=1,\dots,L_1=10$ and $u_{2j}\sim\mbox{Normal}(0,\sigma^2_{u_2})$ for $j=1,\dots,L_2=10$ where the random effects are independent. Then $Y_{ij}\sim\mbox{Normal}(\beta_0+u_{1i}+u_{2j}, \sigma^2)$. Thus the overall sample size is $n=L_1L_2=100$. We take $\sigma^2_{u_1}=0.15$, $\sigma^2_{u_2}=0.10$ and $\sigma^2=0.25$ so the true $R^2\approx 0.46$. 

For R2D2, the full prior specification is
\begin{multline}
       \beta_0\sim\mbox{Normal}(\mu_0,\tau_0^2),\ 
       {\bf u}_1|\phi_1,W\sim\mbox{Normal}(0,\phi_{1} W\bI_{10}),
       {\bf u}_2|\phi_2,W\sim\mbox{Normal}(0,\phi_{2} W\bI_{10}),\\ W|\sigma^2\sim\mbox{GBP}(a,b,1,\sigma^2),\ \bphi\sim\mbox{Dirichlet}(\xi_1,\xi_2),\ \sigma^2\sim\mbox{Inverse-Gammma}(a_0,b_0)
\end{multline}
for hyper-parameters $\mu_0=0,\tau^2_0=100,\xi_1=\xi_2=1$ and $a_0=b_0=0.01$.
Notice that $\sigma^2_{u_1}=\phi_1W$ and $\sigma^2_{u_2}=\phi_2W$.
For the PC prior, the full prior specification is
\begin{multline}
       \beta_0\sim\mbox{Normal}(\mu_0,\tau^2_0),\ {\bf u}_1|\sigma^2_{u_1}\sim\mbox{Normal}(0,\sigma^2_{u_1}\bI_{10}),
       {\bf u}_2|\sigma^2_{u_2}\sim\mbox{Normal}(0,\sigma^2_{u_2}\bI_{10}),\\ \sigma_{u_1},\sigma_{u_2}\sim\mbox{Exp}(\lambda_0),\  \sigma^2\sim\mbox{Inverse-Gammma}(a_0,b_0)
\end{multline}
where $\mu_0=0,\tau^2_0=100,\lambda_0=-\log(0.01)/.968$ and $a_0=b_0=0.01$. The $\lambda_0$ hyperparameter determines the penalty for deviating from the null model where large values of $\lambda_0$ imply a larger penalty. As a default choice, \cite{simpson2017penalising} suggest the value of $\lambda_0=-\log(0.01)/.968$ with interpretation that $P(\sigma_{u_1} > 0.968)=0.01$. This implies (after integrating out $\tau$) a marginal standard deviation for ${\bf u}_1$ and ${\bf u}_2$ of approximately 0.30, which is reasonable for this setting. This choice of hyperparamters yields a prior $R^2$ with a mean of 0.02 and standard deviation of 0.11. The vague prior is the same as the PC prior except $\sigma^2_{u_1},\sigma^2_{u_2}\sim\mbox{InvGamma}(0.5,0.0005)$ \citep{winbugs}, which results in a prior $R^2$ with a mean of $0.22$ and standard deviation of $0.41$.

\begin{table}
\centering
    \begin{tabular}{l|ccccc}
        Prior & $R^2_n$ bias & $R^2_n$ MSE& $Y$ MSE &  $\sigma^2_{u_1}$ MSE & $\sigma^2_{u_2}$ MSE \\\hline
        
        Vague &  -0.06 &0.13 &0.34 &0.14 &0.10 \\
        PC &-0.04 &0.11 &0.33 &0.12 &0.09  \\
        &\vspace{-10pt}\\
        $R^2\sim\mbox{Beta}(1,4)$ &-0.05 &0.11 &0.33 &{\bf 0.09} &{\bf 0.07} \\
        $R^2\sim\mbox{Beta}(1,1)$&-0.02 &0.10 &0.33 &0.12 &0.10 \\
        $R^2\sim\mbox{Beta}(4,1)$ &{\bf 0.01} &{\bf 0.09} &{\bf 0.33} &0.16 &0.13    \\\hline
        S.E. & 0.01 & $<0.01$ & $<0.01$ & 0.01 & $<0.01$
    \end{tabular}
    \caption{Simulation study results for Gaussian regression with random effects and mean$(R^2)=0.46$ and stdev$(R^2)=0.08$. Averaged over 200 repetitions. Largest standard error is in the last row and lowest (absolute) value is in bold.}
    \label{tab:GaussSim1}
\end{table}

The results are in Table \ref{tab:GaussSim1}. The Beta(1,1) and Beta(4,1) priors do the best at estimating $R^2_n$. We can also see that the PC and R2D2 priors are comparable on the holdout $Y$ MSE with the Beta(4,1) prior performing slightly better. The Beta(1,4) prior has a clear advantage estimating the random effects variance. The PC and Beta(1,1) priors are comparable on this metric with the Beta(4,1) and vague priors doing the worst.  The PC prior outperforms the vague prior on all metrics as well as yielding better random effect variance results than the Beta(1,1) and Beta(4,1) prior. Note that the Beta(1,1) prior does not perform the best on every metric, even though its prior mean $R^2$ is closest to the truth. This is likely because of the bias of the sample $R^2$ estimating the population $R^2$ with the random effects in the model (see Appendix A). We also briefly discuss computation time among the different methods. The average number of effective samples per second for the random effect variances is $6500,\ 6300,\ 3100,\ 2600$ and $2600$ for the vague, PCP, $\mbox{Beta}(1,4)$, $\mbox{Beta}(1,1)$ and $\mbox{Beta}(4,1)$, respectively. While the vague and PCP priors are slightly more computationally efficient, all speeds are on the same order of magnitude.

\subsection{Poisson mixed effects model}

We consider a mixed effects scenario for Poisson likelihood as in Section 3.1. Let $\bX_i\sim\mbox{Normal}(0,\Sigma)$ where $\Sigma$ is from a first-order auto-regressive process (AR(1)) with $\rho=0.8$. Let $\beta_0=0.25$ and consider fixed effects $\beta_j\sim\mbox{Normal}(0, 0.1)$ for $j=1,\dots,p=5$. Let there be one random effect $u_j\sim\mbox{Normal}(0,\sigma^2_u)$ for $j=1,\dots,L_1=20$ where all fixed and random effects are independent. Then $Y_{ij}\sim\mbox{Poisson}\{\exp(\beta_0 + \bX_i\bbeta+u_j)\}$ with $i=1,...,m=5$ replicates. Thus the overall sample size is $n=mL_1=100$. We take $\sigma^2_u=0.50$ which gives a true $R^2\approx 0.66$. 

For R2D2, the full prior specification is
\begin{multline}
       \beta_0\sim\mbox{Normal}(\mu_0,\tau_0^2),\
       \bbeta|\phi_1,W\sim\mbox{Normal}(0, \tfrac15\phi_1W\bI_{5}),\
       {\bf u}|\phi_2,W\sim\mbox{Normal}(0,\phi_{2} W\bI_{20}),\\ W\sim\mbox{GBP}(a^*, b^*, c^*, d^*),\ \bphi\sim\mbox{Dirichlet}(\xi_1,\xi_2)
\end{multline}
for hyper-parameters $\mu_0=0,\tau^2_0=3,\xi_1=\xi_2=1$. 

We compare the proposed method with the PC prior. For the PC prior, the full prior specification is
\begin{multline}
       \beta_0\sim\mbox{Normal}(0,\tau_0^2),\        \bbeta\sim\mbox{Normal}(0, \tau_1^2\bI_{5}),\
       {\bf u}|\sigma^2_u\sim\mbox{Normal}(0,\sigma^2_u\bI_{20}),\ \sigma_u \sim\mbox{Exp}(\lambda_0)
\end{multline}
for $\tau_0^2=3,\tau_1^2=100$ and $\lambda_0=-\log(0.01)/.968$. The vague prior is the same as the PC prior except $\sigma^2_u\sim\mbox{InvGamma}(0.5, 0.0005)$. Since the fixed effects have a fixed variance for these two prior specifications, if we consider $\sigma^2_u=W$, then the prior $R^2$ for the vague prior has a mean of $0.46$ and standard deviation of $0.45$. The prior $R^2$ mean and standard deviation for the PC prior is $0.11$ and $0.20$, respectively.


\begin{table}
\centering
    \begin{tabular}{l|ccccc}
        Prior & $R^2_n$ bias & $R^2_n$ MSE & log-score & $||\bbeta-\hat\bbeta||_2$ & $\sigma^2_u$ MSE \\\hline
        Vague & {\bf 0.00} &0.06 &-1.74 &0.57 &0.29 \\
        PC&0.00 &0.06 &-1.74 &0.56 &0.29 \\
        &\vspace{-10pt}\\
        $R^2\sim\mbox{Beta}(1,4)$ &-0.03 &0.07 &-1.72 &{\bf 0.47} &{\bf 0.24}\\
        $R^2\sim\mbox{Beta}(1,1)$  &-0.01 &0.07 &-1.72 &0.48 &0.29 \\
        $R^2\sim\mbox{Beta}(4,1)$ &0.00 &{\bf 0.06} &{\bf -1.72} &0.49 &0.30 \\\hline
        S.E. & $<0.01$ & $<0.01$ & 0.01 & 0.01 & 0.01
    \end{tabular}
    \caption{Simulation study results for Poisson regression with mixed effects and mean$(R^2)=0.66$ and stdev$(R^2)=0.18$. Averaged over 200 repetitions.  Largest standard errors are in last row and lowest (absolute) value is in bold.}
    \label{tab:PoissSim3}
\end{table}

The results are in Table \ref{tab:PoissSim3}. 
The Beta(4,1), PC and vague priors do the best job estimating $R^2_n$. The R2D2 priors give very similar results for log-score and fixed effect estimates with all three of them clearly outperforming the two competing methods. The Beta(1,4) prior again yields the best estimates of the random effect variance but the PC and vague prior do slightly better than the Beta(1,1) and Beta(4,1) R2D2 priors. The Beta(1,4) also does the best at estimating the fixed effects with the other R2D2 priors also outperforming the two competing metrics. Interestingly, the PC prior and vague yield almost identical results across all metrics. Finally, the average number of effective samples per second for the fixed effects is $100,\ 100,\ 190,\ 160$ and $160$, and for the random effect variance is $220,\ 230,\ 240,\ 240$ and $230$ for the vague, PCP, $\mbox{Beta}(1,4)$, $\mbox{Beta}(1,1)$ and $\mbox{Beta}(4,1)$, respectively. All methods have comparable computational efficiency.

\subsection{High-dimensional logistic regression}
Lastly, we consider a logistic regression example with sparsity. Let $n=60$ and $p=50$ and $\bX_i\sim\mbox{Normal}(0,\Sigma)$ where $\Sigma$ is from an AR(1) process with $\rho=0.8$. Let $\beta_0=0.5$ and $\bbeta=(0, {\bf B}_1, 0)$ where ${\bf B}_1\sim\mbox{Normal}(0,1)$ with length 5, i.e., 10\% of the covariates are significant. This makes the true $R^2\approx .37$. 

For R2D2, the full prior specification is
\begin{multline}
       \beta_0\sim\mbox{Normal}(\mu_0,\tau_0^2),\
       \beta_j|\phi_j,W\sim\mbox{Normal}(0, \phi_jW),\\ W\sim\mbox{GBP}(a^*, b^*, c^*, d^*),\ \bphi\sim\mbox{Dirichlet}(\xi_1,\dots,\xi_p)
\end{multline}
for hyper-parameters $\mu_0=0,\tau^2_0=3,\xi_k=1$ for $k\in\{1,\dots,p\}$. For Horseshoe, the full prior specification is
\begin{equation}
       \beta_0\sim\mbox{Normal}(0,\tau_0^2),\        \beta_j|\tau,Z_j\sim\mbox{Normal}(0, Z_j^2\tau^2),\ \tau, Z_1,\dots,Z_p\sim\mbox{Half-Cauchy}(1)
\end{equation} 
where $\tau_0^2=3$. The scale parameter of $1$ for the Half-Cauchy distribution is the default choice given in \cite{carvalho2009handling}. Despite substantial mass near zero for all $\beta_j$, the horseshoe prior also has heavy tails and thus induces a prior distribution on $R^2$ with a mean of $0.92$ and a standard deviation of $0.16$. Lastly, the vague prior takes $\beta_j\sim\mbox{Normal}(0,100)$. Since the fixed effects have a fixed variance, the prior $R^2$ is effectively a point mass at $0.98$.

\begin{table}
    \centering
      \begin{tabular}{l|cccc}
        Prior & $R^2$ bias & $R^2$ MSE & AUC & $||\bbeta-\hat\bbeta||_2$ \\\hline
        Vague & 0.58 &0.58 &{\bf 0.32} &68.65\\
        Horseshoe &0.10 &0.20 &0.28 &8.80  \\
        &\vspace{-10pt}\\
        $R^2\sim\mbox{Beta}(1,4)$&{\bf -0.03} &{\bf 0.15} &0.31 &{\bf 2.64}\\
        $R^2\sim\mbox{Beta}(1,1)$&0.07 & 0.18 &0.31 &5.12 \\ 
        $R^2\sim\mbox{Beta}(4,1)$&0.17 &0.22 &0.30 &7.77 \\\hline
        S.E. & 0.01 & 0.01 & 0.01 & 1.53
    \end{tabular}
    \caption{Simulation study results for Logistic regression with $n=60, p=50$, no random effects and mean$(R^2)=0.35$ and stdev$(R^2)=0.16$. Averaged over 200 repetitions. Largest standard errors are in the last row and lowest (absolute) value is in bold (largest for AUC).}
    \label{tab:LogitSim1}
\end{table}

The results are in Table \ref{tab:LogitSim1}. In this high-dimensional fixed-effects scenario the sample and population definition of $R^2$ are approximately equal (see Appendix A), and thus the Beta(1,4) prior  with mean near the true $R^2$ gives small bias for $R_n^2$. The vague and Horseshoe prior yield a large bias in $R^2$ because their prior $R^2$ has substantial mass near 1 whereas the true $R^2$ is small. The Beta(1,4) and Beta(1,1) priors do the best job estimating $R^2$ which is sensible since their prior mean $R^2$ is close to the true mean $R^2$. Interestingly, the vague prior yields the best AUC. However, estimating the fixed effects is where the R2D2 priors perform particularly well, with the Beta$(1,4)$ performing the best. This is likely attributed to the large prior $R^2$ mass at 0, shrinking the fixed effect estimates towards 0. Lastly, the average number of effective samples per second for the fixed effects is $15,\ 39,\ 120,\ 100$ and $84$ for the vague, Horseshoe, $\mbox{Beta}(1,4)$, $\mbox{Beta}(1,1)$ and $\mbox{Beta}(4,1)$, respectively. Clearly, the R2D2 priors have the greatest computational efficiency for this setting.

Summarizing the results of the simulation study, we find that in most cases the proposed method outperforms current leading approaches. The proposed method has a particular advantage when the true $R^2$ is small and/or when there is sparsity in the fixed effects with the prior inducing $R^2\sim\mbox{Beta}(1,4)$ performing the best. This is likely the case for the sparse example because this prior $R^2$ has a mode at zero which shrinks the parameters to zero. The proposed method also performs well when the true $R^2$ is small and the model has fixed effects because the two competing methods induce a prior on $R^2$ with most of the mass near 1. This is clearly unrealistic in practice and results in a poor model fit. Interestingly, even when the true $R^2$ is large, the $\mbox{Beta}(1,4)$ prior performs the best among the proposed method in terms of estimating the fixed effects and the variance of the random effects.

\section{Real data analysis}

\subsection{Trace plots of Gambia data set}
\begin{figure}
\centering
\subfloat[$\beta_1$]{\includegraphics[width = 2in]{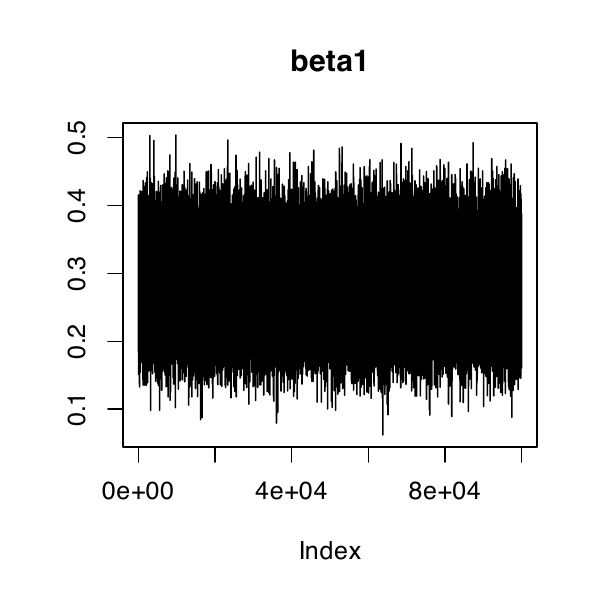}}
\subfloat[$\sigma^2_u$]{\includegraphics[width = 2in]{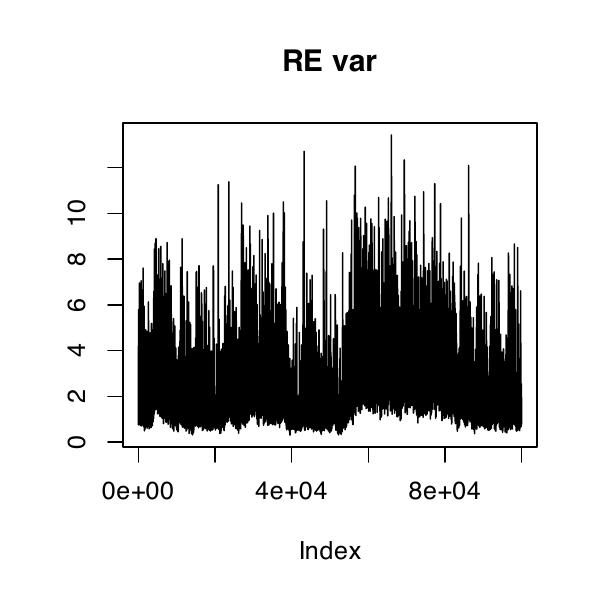}}
\subfloat[$\rho$]{\includegraphics[width = 2in]{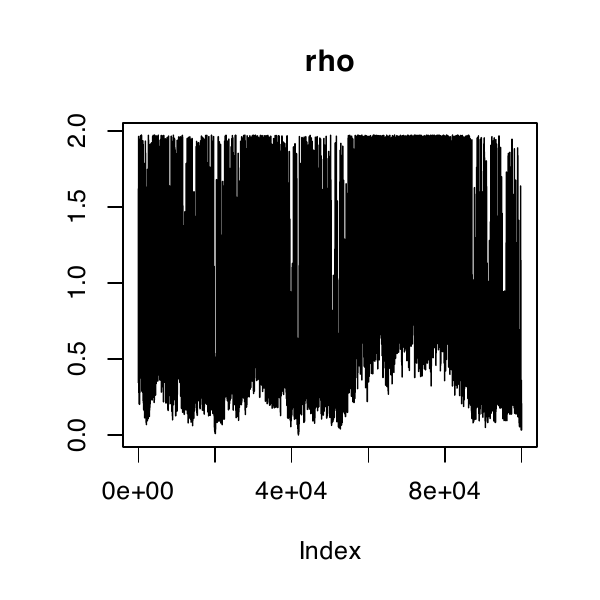}}
\caption{Trace plots for Vague prior on \texttt{Gambia} data set.}
\label{figA:vague}
\end{figure}

\begin{figure}
\centering
\subfloat[$\beta_1$]{\includegraphics[width = 2in]{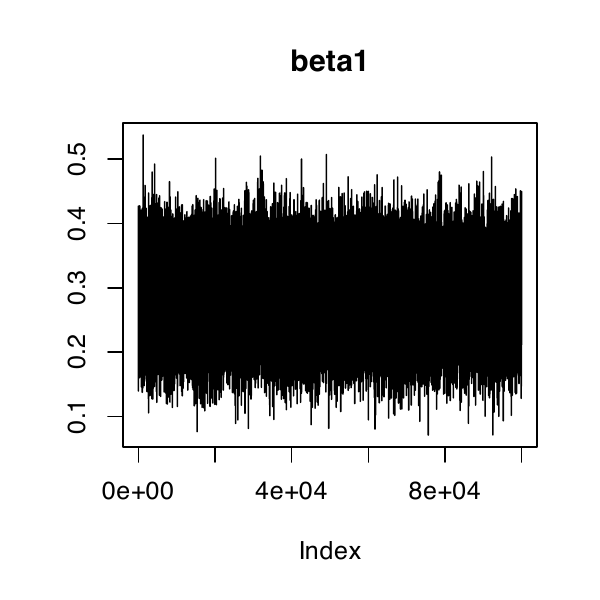}} 
\subfloat[$\sigma^2_u$]{\includegraphics[width = 2in]{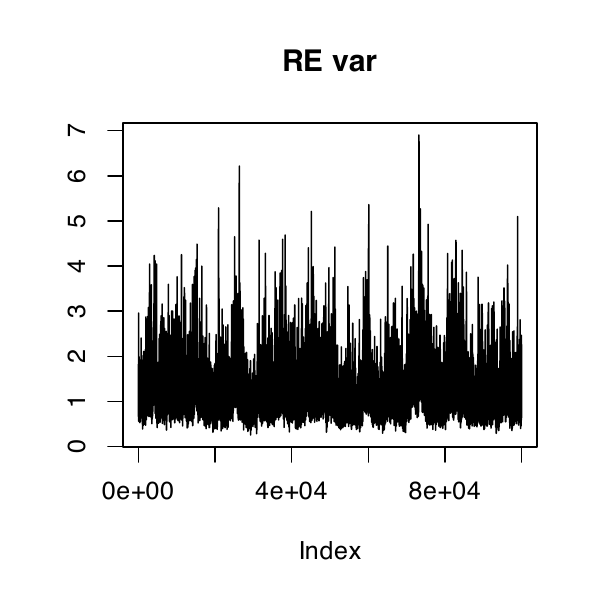}}
\subfloat[$\rho$]{\includegraphics[width = 2in]{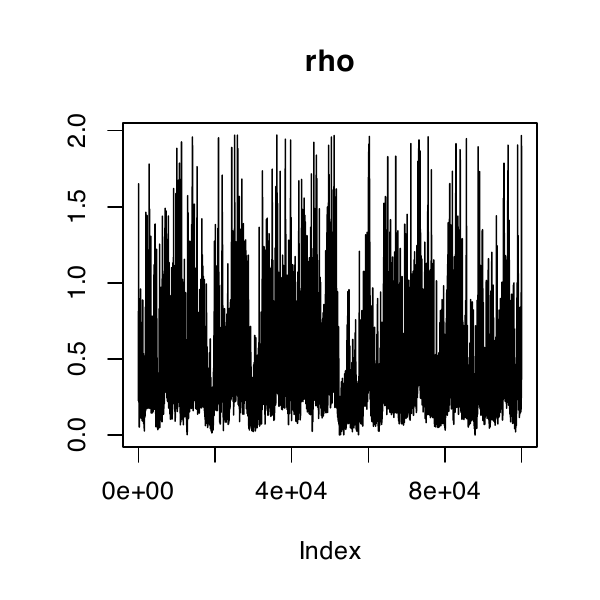}}
\caption{Trace plots for PC prior on \texttt{Gambia} data set.}
\label{figA:pc}
\end{figure}

\begin{figure}
\centering
\subfloat[$\beta_1$]{\includegraphics[width = 2in]{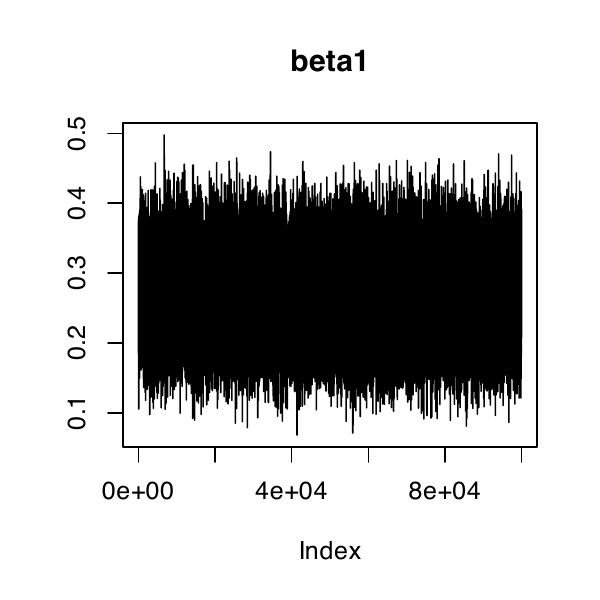}}
\subfloat[$\sigma^2_u$]{\includegraphics[width = 2in]{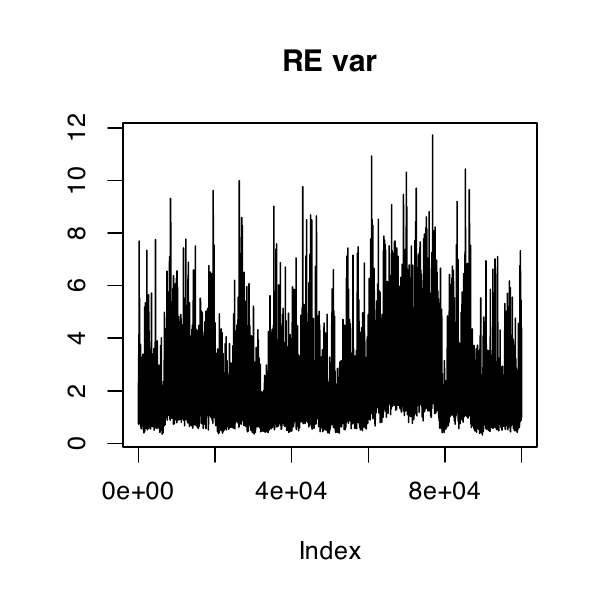}}
\subfloat[$\rho$]{\includegraphics[width = 2in]{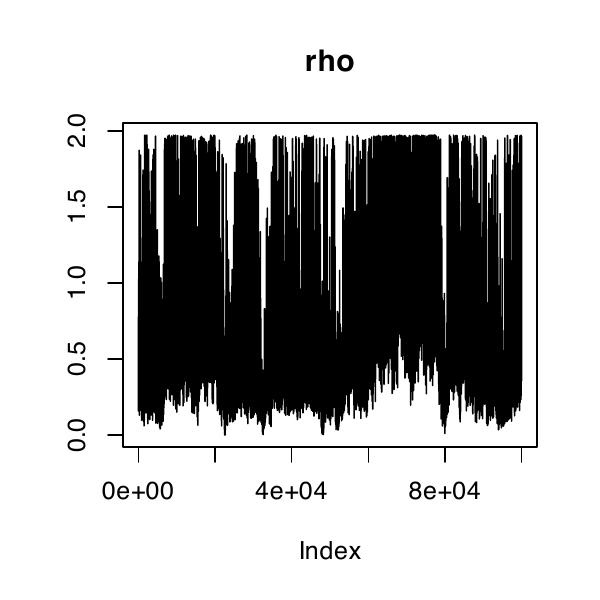}}
\caption{Trace plots for R2D2 prior on \texttt{Gambia} data set.}
\label{figA:r2d2}
\end{figure}

In Figures \ref{figA:vague}-\ref{figA:r2d2} and Table \ref{tab:ess}, we report effective sample sizes (ESS) and trace plots for the Vague, PC and R2D2 priors to check convergence of the MCMC chains. We show the R2D2 prior corresponding to $R^2\sim\mbox{Beta}(1,1)$ as a representative example. All methods have clear convergence for the fixed effect shown $(\beta_1)$. The mixing and ESS are comparable across the different methods for the random effect variance and spatial range parameters. Finally, each method took approximately 20 minutes to draw 100,000 MCMC samples.

\begin{table}[]
    \centering
    \begin{tabular}{c|ccc}
        Method & $\beta_1$ & $\sigma^2_u$ & $\rho$ \\\hline
        Vague & 41,000 & 670 & 640\\
        PC & 36,000 & 1200 & 650\\
        $R^2\sim\mbox{Beta}(1,1)$ & 28,000 & 800 & 620
    \end{tabular}
    \caption{Effective sample sizes for various parameters in the \texttt{Gambia} data analysis.}
    \label{tab:ess}
\end{table}

\subsection{Genomics data results}
In Table \ref{tab:gen}, we report the average effective sample size for $\bbeta$ and the global variance parameter, as well as the computing time.

\begin{table}[h]
    \centering
    \begin{tabular}{l|ccc}
        Method & ESS $\bbeta$ & ESS var & Time  \\\hline
        Horseshoe & 9900 & 9800 & 2500 \\
        $R^2\sim\mbox{Beta}(1,5)$ & 9900 & 9800 & 1000  \\
        $R^2\sim\mbox{Beta}(1,10)$& 9800 & 9800 & 1500 \\
        $R^2\sim\mbox{Beta}(1,20)$& 9800 & 9800 & 1000\\
        $R^2\sim\mbox{Beta}(1,30)$& 9800 & 9800 & 1000
    \end{tabular}
    \caption{Results for genomics data analysis. Effective sample size for $\bbeta$ and global variance term ($\tau^2$ for Horseshoe and $W$ for R2D2), and computing time in seconds.}
    \label{tab:gen}
\end{table}

\end{document}